\documentclass[useAMS, usenatbib]{mn2e}

\usepackage[dvips]{graphicx,color}
\usepackage[]{txfonts}
\usepackage[svgnames]{xcolor}
\usepackage{soul}

\def\simless{\mathbin{\lower 3pt\hbox
{$\rlap{\raise 5pt\hbox{$\char'074$}}\mathchar"7218$}}}   
\def\simmore{\mathbin{\lower 3pt\hbox
{$\rlap{\raise 5pt\hbox{$\char'076$}}\mathchar"7218$}}}   

\def\difd{\mathrm{d}}
\newcommand{\dsfrac}[2]{\displaystyle{\frac{#1}{#2}}}

\newcommand{\secref}[1]{$\S$~\ref{#1}}
\newcommand{\figref}[1]{Fig.~\ref{#1}}
\usepackage{enumerate}

\newcommand{\ubs}{Swift\,\,J1644+57}

\newcommand{\corr}[1]{{#1}}

\title[Swift J1644+57: powerful jet with fast core and slow sheath]{The radio afterglow of Swift J1644+57 reveals a powerful jet with fast core and slow sheath}

\author[P. Mimica, D. Giannios, B. D. Metzger and M. A. Aloy]
{P. Mimica$^{1}$\thanks{E-mail: Petar.Mimica@uv.es}, D. Giannios$^{2}$, B. D. Metzger$^{3}$ and M. A. Aloy$^{1}$\\
$^{1}$Departamento de Astronom\'ia y Astrof\'isica, Universidad de Valencia, 46100, Burjassot, Spain\\
$^{2}$Department of Physics and Astronomy, Purdue University, 525 Northwestern Avenue, West Lafayette, IN 47907, USA\\
$^{3}$Columbia Astrophysics Laboratory, Columbia University, New York, NY, 10027, USA}

\begin{document}

\maketitle

\label{firstpage}

\begin{abstract}

We model the non-thermal transient \ubs~as resulting from a relativistic jet powered by the accretion of a tidally-disrupted star onto a super-massive black hole.  Accompanying synchrotron radio emission is produced by the shock interaction between the jet and the dense circumnuclear medium, similar to a gamma-ray burst afterglow.  An open mystery, however, is the origin of the late-time radio re-brightening, which occurred well after the peak of the jetted X-ray emission.  Here, we systematically explore several proposed explanations for this behaviour by means of multi-dimensional hydrodynamic simulations coupled to a self-consistent radiative transfer calculation of the synchrotron emission.   Our main conclusion is that the radio afterglow of \ubs~is not naturally explained by a jet with a one-dimensional top-hat angular structure.  However, a more complex angular structure comprised of an ultra-relativistic core (Lorentz factor $\Gamma \sim 10$) surrounded by a slower ($\Gamma \sim $ 2) sheath provides a reasonable fit to the data.  Such a geometry could result from the radial structure of the super-Eddington accretion flow or as the result of jet precession. The total kinetic energy of the ejecta that we infer of $\sim$ few $10^{53}\,$erg requires a highly efficient jet launching mechanism.  Our jet model providing the best fit to the light curve of the on-axis event \ubs~is used to predict the radio light curves for off-axis viewing angles.  Implications for the presence of relativistic jets from TDEs detected via their thermal disk emission,  as well as the prospects for detecting orphan TDE afterglows with upcoming wide-field radio surveys and resolving the jet structure with long baseline interferometry, are discussed.  
\end{abstract}

\begin{keywords}
radiation mechanisms: non-thermal -- galaxies: active – galaxies: jets – galaxies: nuclei -- hydrodynamics -- radiative transfer
\end{keywords}

\section{Introduction}
\label{sec:intro}

Radio astronomy is undergoing a revolution in the study of time-domain phenomena due to the advent of sensitive wide-field arrays.  At meter wavelengths these include LOw Frequency ARray (LOFAR; \citealt{vanHaarlem+13}), the Murchison Widefield Array (MWA; \citealt{Lonsdale:2009aa}) and the Long Wavelength Array (LWA; \citealt{Ellingson:2009aa}).  At centimetre wavelengths (GHz frequencies) a new generation of wide-field facilities is also being developed, such as Apertif/Westerbork Synthesis Radio Telescope (WSRT; \citealt{Oosterloo:2009aa}), MeerKAT \citep{Booth:2009aa}, and the Australian Square Kilometer Array Pathfinder (ASKAP; \citealt{Johnston:2008aa}).  However, despite the overall maturity of radio astronomy, surprisingly little is known about what astrophysical sources will dominate the transient sky.  

The brightest sources at meter-wavelengths on short timescales are expected to result from coherent processes such as `giant pulses' from Galactic pulsars (e.g.~\citealt{Jessner+05}) and cyclotron maser emission from brown dwarfs and planets \citep{Berger:2002aa,Hallinan:2008aa}.  Brighter extra-galactic analogs of such events have not yet been detected \citep[although see][]{Thornton:2013aa,Spitler:2014aa}, but may accompany rare violent events, such as giant magnetar flares (\citealt{Lyubarsky:2014aa}) and the merger of binary neutron stars \citep{Hansen:2001aa}.  

At centimeter wavelengths and on longer timescales, incoherent synchrotron sources may instead dominate the transient sky.  These generally result from the shock interaction between rapidly expanding matter from an energetic explosion and dense ambient gas.  Known examples include radio supernovae \citep{Weiler:2002aa}, off-axis (`orphan') afterglows of gamma-ray bursts \citep{Totani:2002aa}, and other relativistic outflows from the core collapse of massive stars \citep[e.g., SN 2009bb; ][]{Soderberg:2010aa}.  The transient {\it Swift}  J164449.3+573451 (hereafter \ubs) is the prototype for a new type of relativistic transient that will also contribute significantly to the variable radio sky (\citealt{Giannios:2011aa}).  

\ubs~was characterised by powerful non-thermal X-ray emission (\citealt{Bloom:2011aa}; \citealt{Levan:2011aa}; \citealt{Burrows:2011aa}; \citealt{Zauderer:2011aa}).  The long duration of \ubs~and a position coincident with the nucleus of a previously quiescent galaxy lead to the conclusion that it was powered by rapid accretion onto the central supermassive black hole (SMBH) following the tidal disruption of a star (a `tidal disruption event', or TDE; \citealt{Carter:1982aa,Rees:1988aa}), although alternative explanations have been proposed (e.g.~\citealt{Quataert&Kasen12}).  The rapid X-ray variability suggested an origin internal to a jet that was relativistically beaming its radiation along our line of sight, similar to the blazar geometry of normal active galactic nuclei \citep{Bloom:2011aa}.  \ubs~was also characterised by luminous synchrotron radio emission, that brightened gradually over the course of several months (\citealt{Zauderer:2011aa}).  Unlike the rapidly varying X-ray emission, the radio emission resulted from the shock interaction between the TDE jet and the dense external gas surrounding the SMBH (\citealt{Giannios:2011aa}; \citealt{Bloom:2011aa}), similar to GRB afterglows.  A second jetted TDE, Swift J2058+05, with similar  X-ray and radio properties to \ubs~was reported by \citet{Cenko:2012aa}.

Jetted TDEs in principle offer a unique opportunity to witness the birth of an AGN, thus providing a natural laboratory to study the physics of jet production across a wide range of mass feeding rates.  Obtaining a better understanding of \ubs~and J2058+05 would thus have far-reaching consequences for topics such as the physics of relativistic jet formation and super Eddington accretion, the conditions (e.g.~distribution of accreting or outflowing gas) in nominally quiescent galactic nuclei, and possibly even the astrophysical origin of ultra-high energy cosmic rays (\citealt{Farrar&Gruzinov09}; \citealt{Farrar&Piran15}).  Despite this promise, theory has yet to fully exploit the wealth of data available for \ubs.  Accurate, multi-dimensional models for TDE jets are needed to quantify how similar events would appear to observers off the jet axis.  Such information is necessary to constrain the presence of off-axis jets in TDEs detected via their quasi-isotropic thermal emission (\citealt{Bower:2013aa}; \citealt{vanVelzen:2013aa}) or to assess how jetted TDEs will contribute to future radio transient surveys (\citealt{Giannios:2011aa}; \citealt{Frail:2012aa}). 

Perhaps the greatest mystery associated with \ubs~is the unexpected flattening or rebrightening of the high frequency radio light curves on a timescale of a few months after its initial rise (\citealt{Berger:2012aa}; \figref{fig:radio}).  This behavior contradicts expectations of one-dimensional blast wave models that assume energy input proportional to the observed X-ray emission, and which instead predict that the high frequency emission should decay rapidly once the jet energy is transfered to the blast (\citealt{Metzger:2012aa}).  This behavior suggests that either the temporal or angular structure of the jet is more complex than assumed in the basic models usually applied to GRB jets (e.g., \citealt{Berger:2012aa}; \citealt{Wang:2014aa}), or that some key physical process has been neglected (e.g., \citealt{Kumar:2013aa}).  

In this paper we model the hydrodynamical interaction between a TDE jet and the circumnuclear medium (CNM) using one and two-dimensional relativistic hydrodynamical simulations.  Assuming that non-thermal electrons are accelerated at the shock fronts, we exploit the simulation results to perform a radiative transfer calculation for the synchrotron radio emission.  Our goal is to provide a comprehensive study of the structure of the jet that gave rise to \ubs, in order to identify which (if any) of the proposed models best reproduce the radio data.  

This paper is organised as follows.  In $\S\ref{sec:obs}$ we review the observations of \ubs~and summarise previous work on its interpretation. \secref{sec:model} describes our model for the jet and the technical details of scenarios that we consider.  In $\S\ref{sec:results}$ we describe our results.  \secref{sec:app} describes applications and implications of our results, including the inferred physical structure of TDE jets powered by super-Eddington accretion and the prospects for detecting orphan jetted TDEs with future radio surveys.  In $\S\ref{sec:concl}$ we summarise our results.

\section{A Jetted Tidal Disruption Event}

\label{sec:obs}

\subsection{Observations of \ubs}

\begin{figure}
  \centering
  \includegraphics[width=8.6cm]{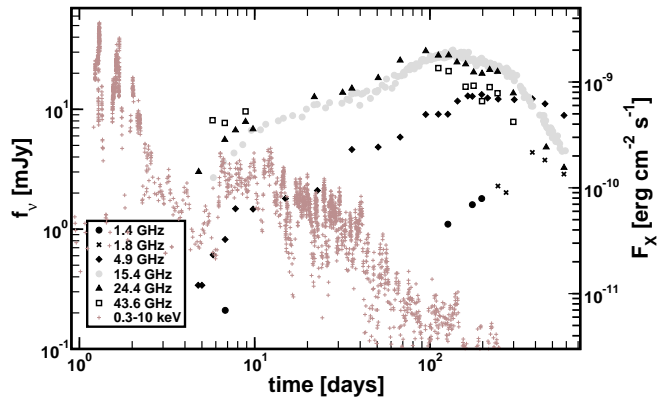}
  \caption{Radio light curves of \ubs~at observing frequencies $1.4$, $1.8$, $4.9$, $15.4$, $24.4$
    and $43.6$ GHz from \citet{Berger:2012aa} and \citet{Zauderer:2012aa}. 
Also shown are the \emph{Swift} XRT observations in the $0.3-3$ keV band (brown crosses, right axis).
Note the broad secondary maximum in the radio occurring $\sim 6$ months after trigger, well after the X-ray maximum.}
  \label{fig:radio} 
\end{figure}

The two distinct components in the SED of \ubs~indicated different origins for the X-ray and radio emission (\citealt{Bloom:2011aa}, \citealt{Burrows:2011aa}).  Rapid variability of the X-rays placed their origin at small radii close to the SMBH, likely from a source internal to the jet itself.  Constraints on the brightness temperature and variability of the radio emission instead place its origin at larger radii, where the relativistic outflow began to interact with the CNM surrounding the black hole (\citealt{Giannios:2011aa}). Figure \ref{fig:radio} shows the observed radio and X-ray light curves of \ubs.

After several days of peak activity, the X-ray luminosity decreased as a power law $L_{x} \propto t^{-\alpha}$ in time with $\alpha \sim 5/3$, consistent with the predicted decline in the fall-back rate $\dot{M}$ of the disrupted star (\citealt{Rees:1988aa}). Unless coincidental, the apparent fact that $L_x \propto \dot M$ implies that both the jet power $L_j$ tracked the instantaneous accretion rate and that the radiative efficiency and the bulk Lorentz factor of the outflow remained relatively constant throughout the event\footnote {Substantial temporal variation of the bulk Lorentz factor would cause equally great changes in the beaming degree of the jet emission, making hard to understand why $L_x \propto \dot M$
for $\sim 1$ year after the TDE.}.  

A model for the radio emission of \ubs~using the first three weeks of data was developed by \citet[][hereafter MGM12]{Metzger:2012aa}, under the assumption that the X-ray luminosity tracked the instantaneous kinetic power of the jet.  This sudden injection of energy into the CNM produces a forward-reverse shock structure, which acts to decelerate the ejecta.  The evolution is characterised by two stages: (1) an early phase during which the reverse shock crosses back through the relativistically expanding ejecta that was released over the initial few-day period of peak activity; (2) a later phase during which the relativistic blast wave undergoes self-similar expansion \citep{Blandford:1976aa}.

A striking feature of the early-time radio light curve was the presence of an achromatic break at $t \sim 10$ days (\figref{fig:radio}).  MGM12 showed that this break could result from the transition between stage (1) and (2), and that the temporal slopes of the pre- and post-break light curve requires that the CNM density $n_{\rm cnm}$ scale with radius as $\propto r^{-\beta}$ with $\beta \sim 2$.  MGM12 also constrained the initial (unshocked) Lorentz factor $\Gamma_{\rm j} \sim 10$ and opening angle $\theta_{\rm j} \sim 1/\Gamma_{j} \sim 0.1$ to values remarkably similar to those of blazar jets (e.g.~\citealt{Pushkarev:2009aa}). The estimated value of the CNM density $\sim$ few cm$^{-3}$ at radii $\sim 10^{18}$ cm from the SMBH was somewhat lower than that measured on a similar radial scale from SgA* in our own galaxy (\citealt{Berger:2012aa}).  The reverse shock indirectly shaped the early light curves of \ubs~through its effect on the hydrodynamics, but the observed synchrotron emission was probably dominated by electrons accelerated at the forward shock (MGM12)\footnote{With the possible exception of the radio emission during the first $\sim$10 days after trigger; see below.}. 

\citet{Berger:2012aa} presented updated radio light curves of \ubs, which showed the surprising radio rebrightening mentioned earlier (\figref{fig:radio}). This behavior deviated significantly from that predicted by MGM12 for a blast-wave evolving with constant energy, thus demanding that an additional source of energy contribute to the forward shock at late times (e.g.~\citealt{Barniol:2013aa}).  Although a temporary rebrightening can be created by the self-absorption frequency passing through the observers frequency band (MGM12), such a transition cannot explain the observed light curves. 

\subsection{Proposed explanations for the radio rebrightening}
\label{sec:explanations}

After the one-zone jet model failed to reproduce the late radio excess observed for \ubs, several authors proposed explanations for this apparent increase in the energy of the forward shock \citep{Barniol:2013aa}.  We briefly review these possibilities here.  

\citet{Berger:2012aa} proposed that the radio rebrightening can be explained as the result of slower material catching up and injecting energy to the decelerating forward shock at late times (e.g., \citealt{Granot&Kumar06}).  In this model the slower material is ejected after the initial ultra relativistic jet stage and it contains $\sim$ 20 times more energy than the $\Gamma_{\rm j}\sim 10$ ejecta. The decline of the X-ray lightcurve was believed to reflect the decrease in the jet power corresponding to the predicted evolution of the rate of mass fallback to the SMBH, indicating that the beaming of the jet emission (and therefore the jet Lorentz factor) did not vary substantially over that period. It is thus not clear in this scenario what process revived the source, resulting in the required slower and more powerful second ejection event. On the other hand, the energetic, slower ejecta may have been ejected simultaneously with the relativistic component, e.g., encasing it (see below).   

\citet{DeColle:2012aa} performed 2D hydrodynamic simulations of the interaction between the TDE jet and its surrounding environment.  The latter was assumed to include both the extended hydrostatic atmosphere which is produced by returning stellar debris (\citealt{LoebUlmer97}; \citealt{Guillochon+14}) on small radial scales ($\lesssim 10^{-3}$ pc) and the CNM of the nuclear cluster on larger scales.  The cocoon created by shocked CNM plays an important role in the model of \citet{DeColle:2012aa} in collimating the jet and, potentially, impacting the late-time radio emission.  The jet is only self-collimated by its environment, however, if its kinetic luminosity is less than a critical value which we estimate to be $\sim 10^{44}$ erg s$^{-1}$ using the formalism of \citet{Bromberg+11}, i.e.~well below the peak (beaming-corrected) luminosity of \ubs, $\gtrsim 10^{45-46}$ erg s$^{-1}$.

Radio rebrightening could in principle also result from the jet encountering an abrupt change in the radial slope of the CNM density profile.  Such a break might be expected because on small radial scales the density profile is expected to approach that of a Bondi accretion flow onto the SMBH ($n_{\rm cnm} \sim r^{-3/2}$), while on larger scales outside the SMBH sphere of influence the density profile should flatten.  However, past studies of GRB afterglows show that changes in the external density profile produce only minor changes in the synchrotron light curve behavior \citep[e.g.][and references therein]{Mimica&Giannios11,Gat:2013aa} and rarely rebrightening.  

\citet{Kumar:2013aa} emphasise that electrons accelerated at the forward shock are subject to Compton cooling by the X-rays from the jet.  According to their model, the apparent excess in the late radio emission is in fact a {\it deficit} in the early-time emission, resulting from the synchrotron flux being suppressed due to the lower effective cooling frequency of X-ray cooled electrons.  Although X-ray cooling will certainly play a role in shaping the early radio emission, it becomes less efficient as the jet spreads laterally, since a smaller fraction of the freshly accelerated electrons (only those within the opening angle of the original jet) are cooled.  \citet{Kumar:2013aa} also consider the effect of X-ray cooling on optically thin emission. As we show here, its effect is negligible for the radio bands at which the observed emission is self absorbed. 

A final possibility, which we advocate in this work, is that the late radio brightening indicates that the jet responsible for \ubs~possessed a more complex angular structure than the top hat structure adopted in most previous models.  A natural alternative is that the jet possesses an ultra-relativistic ($\Gamma_j \sim 10$) core responsible for the hard X-ray emission, which is surrounded by a wider-angle, mildly relativistic ($\Gamma_j \sim 2$) sheath.  Core emission dominates the early-time emission, while the more energetic sheath dominates the late radio emission (e.g.~\citealt{Tchekhovskoy:2014aa}; \citealt{Wang:2014aa}; see, however, \citealt{Liu:2015aa}).  This scenario allows both components of the jet to be ejected simultaneously during the first weeks following the TDE, but the emission is initially dominated by that produced at the external shocks driven by the fast core.  Such a jet structure is physically motivated if the slow ejecta results from an initial phase of precession of the fast jet (\citealt{StoneLoeb12}; \citealt{Tchekhovskoy:2014aa}) or simply as the result of the separate fast and slow outflow components expected to accompany super-Eddington accretion (\citealt{Mineshige:2005aa}; \citealt{Coughlin:2014aa}; \citealt{Jiang+14}).  A similar fast core/slow sheath configuration is motivated by blazar observations (\citealt{Ghisellini05}), indicating that such a structure may be a generic feature of relativistic jets.

The rich data set available for \ubs, now spanning over three years, makes it possible to construct a much more detailed model for the afterglow, which explores the effects of late energy injection, complex angular structure, and X-ray cooling of the radiating electrons described above.  However, most models to date include only one dimensional hydrodynamics.  Multi-dimensional effects, such as angular structure and jet spreading or pressure confinement by the shocked external medium, become important as the jet decelerates to mildly relativistic speeds.  Indeed, since by now, $\sim 4$ years after the trigger, even the fastest components of the jet have decelerated to $\Gamma\lesssim 2$, modeling the jet as part of a spherical flow is clearly inadequate.  Multidimensional models are also necessary to properly address the emission observed by off-axis viewers, which peak soon after the jet becomes mildly relativistic.  Below we describe a series of one and two-dimensional hydrodynamic simulations that explore systematically the above physical effects in order to more accurately model the radio afterglow of \ubs.  

\section{Model Description and Numerical Procedure}
\label{sec:model}

The numerical code \verb'MRGENESIS' \citep{Mimica:2007aa,Mimica:2009bb} is used to perform 1D and 2D relativistic hydrodynamic simulations in spherical symmetry. \verb'MRGENESIS' is a high-resolution, shock-capturing scheme which uses MPI/OpenMP for hybrid parallelization and HDF5 libraries for parallel input and output.

The TDE jet responsible for \ubs~is modeled as a relativistic outflow with a temporally constant Lorentz factor $\Gamma_{\rm j}$ and a kinetic luminosity $L_{\rm j}$ that varies in time with a form motivated by the observed (isotropic) X-ray light curve,
\begin{equation}\label{eq:lum}
  L_{\rm j}(t) = \left\{\begin{array}{rl}
      &L_{\rm j, 0} \,\,\,\,\,\mathrm{if}\,\,\,\,\, t\leq t_{\rm j}\\[4mm]
      &L_{\rm j, 0} \left(\dsfrac{t}{t_{\rm j}}\right)^{-5/3} \,\,\,\,\,\mathrm{if}\,\,\,\,\, t>t_{\rm j},
    \end{array} \right.
\end{equation}
where $t_{\rm j} = 5\times 10^{5}$ s $\sim$1 week is the approximate duration of the epoch of peak X-ray activity (e.g., \citealt{Burrows:2011aa}).  The peak luminosity $L_{\rm j,0}$ is related to the total isotropic energy of the jet according to $E_{\rm iso} = \int_0^{\infty} L_j(t)\ {\rm d}t \approx 5L_{\rm j,0}t_{\rm j}/2$.  Motivated by the isotropic energy radiated in the soft X-ray band $\sim 3\times 10^{53}$erg (\citealt{Levan:2011aa}; \citealt{Burrows:2011aa}),
we require that $E_{\rm iso}\gtrsim  3\times 10^{53}$erg.  The true (beaming-corrected) energy of the jet is $E = E_{\rm iso}f_{\rm b}^{-1}$, where {$f_{\rm fb} = (1-\cos{ \theta_{\rm j}})$} is the beaming fraction.  In our model for fast, narrow jets we assume an initial jet Lorentz factor $\Gamma_{\rm j} = 10$ and jet half-opening angle $\theta_{\rm j} = 1/\Gamma_{\rm j} = 0.1$.  We also consider lower values of $\Gamma_{\rm j,s} = 2$ for $\theta < \theta_{\rm j,s} = 1/\Gamma_{\rm j,s} = 0.5$ in the case of a wider, slower jet, or two-zone models with both fast and slow components.  

The inner radius of the numerical grid, $R = r_{\rm j,0}$, corresponds to the location where the jet is injected.   The initial size of the radial grid for 1D simulations is $r_{\rm out,0}=3.2 \times 10^{17}\,$cm, but this is periodically enlarged as the jet propagates to larger radii. This enlargement occurs in such a way as to keep the numerical resolution per unit length constant in the radial direction, i.e. new uniform zones are added to the grid each time it is extended.  For 2D simulations the angular coordinate $\theta$ has a fixed range $[0, \pi/2]$, while the jet is injected at $R = r_{\rm j, 0}$ across the angular range $\theta \leq \theta_{\rm j}$.  The ratio of the jet pressure to rest mass density is $P_{\rm j}(t)/\rho_{\rm j}(t) c^2=0.01$, so that it is sufficiently small for the initial thermal pressure of the jet material to be negligible and, hence, the specific jet enthalpy, $h_{\rm j}=1+\varepsilon_{\rm j}/c^2+P_{\rm j}(t)/\rho_{\rm j}(t) c^2\simeq 1$ (where $\varepsilon_{\rm j}$ is the specific internal jet energy).  The jet density $\rho_{\rm j}(t)$ at the inner boundary is then related to the jet luminosity and Lorentz factor according to
\begin{eqnarray}
\rho_j(t) = \frac{L_{\rm j}(t)}{4\pi r_{\rm j,0}^{2}h_{\rm j} \Gamma_{\rm j}^{2}v_{\rm j}c^2},
\label{eq:rhoj}
\end{eqnarray}
where $v_{\rm j}$ is the jet velocity.
The external CNM is also initialised to a cold temperature, with a power-law radial density profile $n_{\rm cnm} \propto r^{-k}$ with $k \sim 1-2$. \corr{Finally, the jet to external medium effective inertia density ratio $\eta_R=\rho_jh_j\Gamma^2_j/(\rho_{\rm cnm}h_{\rm cnm}\Gamma_{\rm cnm}^2)$ is set to be constant in all our simulations and roughly equal to 1000.}

\subsection{1D Simulation Details}
\label{sec:1Ddetails}

A numerical grid with radial resolution $\Delta r = 1.1\times 10^{13}$ cm extends from the inner boundary $R = r_{j, 0} = 10^{16}$ cm. The latter value is taken to be sufficiently small so that our simulations adequately follow the initial jet-CNM interactions including the reverse shock crossing of the jet. The jet evolution is followed for a total duration of $\sim 6.7\times 10^8$\,s .  A total of $1600$ snapshots of the jet state are saved at regular intervals (every $8.3\times 10^4$\,s) during the first $1.3\times 10^8$\,s, afterward the time interval between the snapshots is progressively increased, resulting in an additional $785$ snapshots being stored over the remaining $\sim 5.4\times 10^8$\,s. The collection of saved snapshots is used as an input for the \verb'SPEV' code \citep{Mimica:2009aa} in order to compute the synchrotron radio light curves.  Appendix~\ref{app:conv} provides additional technical details on the light curve calculation in \verb'SPEV'.

\subsection{2D simulation details}
\label{sec:2Ddetails}

2D models are simulated using the same initial and boundary conditions as in the equivalent 1D case. The radial and angular resolution are $\Delta r = 4\times 10^{14}$ cm and $\Delta\theta = \pi/600$ rad, respectively. We output the snapshots at exactly the same evolutionary times as in 1D case (\secref{sec:1Ddetails}). It should be remarked that $\Delta r$ and $\Delta\theta$ are chosen to make the \verb'SPEV' calculations feasible, since \verb'SPEV' must solve a '3+1'-dimensional ray-tracing problem, i.e. the main contributor to the computational complexity of this work is \verb'SPEV', not \verb'MRGENESIS'. 

\subsection{Non-thermal particles and emission}
\label{sec:nonthermal}

Energetic electrons are injected behind the forward shock with an energy spectrum $f(\gamma) \propto \gamma^{-p}$, as meant to mimic the effects of e.g. diffusive shock acceleration.  The forward shock is identified using the algorithm described in Sec. 3.1. of \citet{CuestaMartinez:2014wq}.  Particle acceleration at the RS is ignored, motivated by the lack of evidence for a substantial RS emission component from the observed radio light curves of \ubs~(MGM12; although the RS may be contributing to the low-frequency emission during the first week post trigger--see below).  Fractions $\epsilon_e$ and $\epsilon_B$ of the post-shock thermal energy are placed into relativistic electrons and the magnetic field, respectively. \corr{Optionally, we allow that only a fraction $\zeta_e$ of available electrons is accelerated at shocks (we adopt $\zeta_e=1$ except where indicated in Section~\ref{sec:1D2Z} and in Appendix~\ref{app:zeta_e}).}  We adopt $p=2.2$, $\epsilon_e = 0.1$ and $\epsilon_B = 10^{-3}$ as fiducial values of the microphysical parameters, similar to those used to fit GRB afterglows \citep[e.g.,][]{Mimica:2010aa}.

Injected particles lose energy to synchrotron cooling and adiabatic losses, and in some cases we also include inverse Compton (IC) cooling by X-ray photons (IC cooling occurs in the Thomson regime for parameters of interest). The electron Lorentz factors $\gamma$ evolve according to
\begin{equation}\label{eq:cool}
\dsfrac{\difd\gamma}{\difd t'} = \dsfrac{1}{3}\dsfrac{\difd\ln\rho}{\difd t'}\gamma - \dsfrac{4}{3}\dsfrac{c\sigma_T}{m_e c^2}\left(u'_B + u'_{\rm rad}\right)\gamma^2
\end{equation}
where $t'$, $\rho$, $u'_B$ and $u'_{\rm rad}$ are the time, fluid density, magnetic field and external radiation energy density, respectively, in the comoving frame of the fluid.  The value of $u'_{\rm rad}$ is only nonzero for models including IC cooling (\secref{sec:Xraycool}).

\section{Results}
\label{sec:results}

Exploring each of the proposed scenarios for the afterglow of \ubs~($\S\ref{sec:explanations}$) across their full parameter space would require at least tens of numerical simulations, resulting in a non-thermal emission computational time that is currently not feasible for two dimensional (axisymmetric) simulations.  We thus adopt the strategy of first performing a large number of exploratory runs under the assumption of spherical symmetry for the blast wave evolution ($\S\ref{sec:1D1Z}$, $\S\ref{sec:1D2Z}$).  Resulting 1D light curves are then compared to the observations, allowing us to identify the most promising models, along with the jet/CNM parameters that best reproduce the observations.  These most promising scenarios are then explored in great detail by repeating them using a full two dimensional calculation ($\S\ref{sec:2D}$).  
  
\subsection{One-component 1D models}\label{sec:1D1Z}

We begin by considering light curves produced by \corr{one-component,} spherically symmetric jets with an assumed (fixed) opening angle $\theta_{\rm j}$.  Note that although the initial jet opening angle $\theta_{\rm j}$ does not factor into the hydrodynamic evolution in the one dimensional model, its value must still be specified because the angular size of the emitting region does factor into the emission calculation\footnote{Contrast this with GRB afterglows, in which the shocked jet material has $\Gamma_{\rm sh}\theta_{j}>1$ prior to the jet break, i.e., the finite opening angle of the jet does not affect the observed flux.  However, for TDE jets, $\Gamma_{\rm sh}\theta_{j}\simless 1$, thereby suppressing the observed flux by a factor of $\sim (\Gamma_{\rm sh}\theta_{j})^2$ with respect to that from an otherwise equivalent spherical outflow \citep{Bloom:2011aa}. Models for \ubs~that assume a spherically symmetric flow may strongly overestimate the afterglow emission of the actual beamed event.}.  

\subsubsection{Fast, narrow jet}
\label{sec:FNjet}

We first consider a fast, narrow jet with an isotropic energy $E_{\rm iso} = 10^{53}$ erg expanding into a stratified external medium with the radial density profile $n_{\rm cnm} = n_{18}(r/10^{18}{\rm cm})^{-k}$, where $n_{18} = 1.5$ cm$^{-3}$ and $k$ is varied between 1, 3/2, and 2.  The jet Lorentz factor and half-opening angle are assumed to be $\Gamma_{j}=10$ and $\theta_{j}=0.1$, respectively.  

\begin{figure}
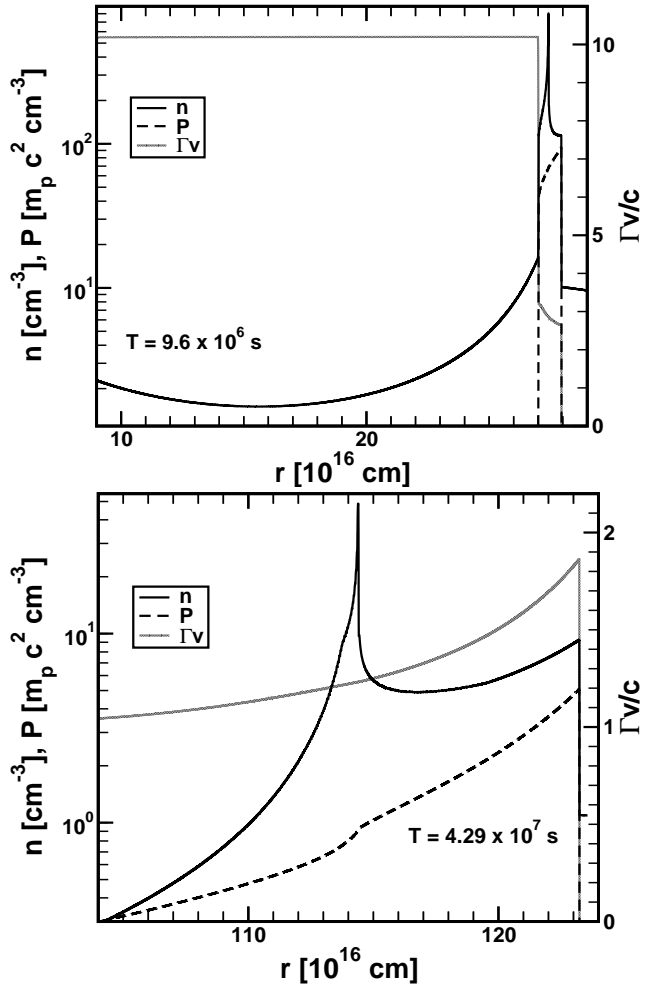

  \centering
  \includegraphics[width=8.4cm]{figures/hydro1D-early.eps}
  \includegraphics[width=8.4cm]{figures/hydro1D-late.eps}
  \caption{\emph{Upper panel:} Number density $n$ ({\it \corr{full line}}), thermal pressure $P$ ({\it \corr{dashed line}}) and four-velocity $\Gamma v/c$ ({\it \corr{grey line}}) of the fast, narrow jet at the time $T \sim 9.6\times 10^6$ s.  The reverse and forward shocks are visible as discontinuities on the right half of the plot.  The density peak marks the location of the contact discontinuity in both the top and bottom panels.  \emph{Lower panel:} Same as upper panel, but for a later stage in the jet evolution ($T\sim 4.3\times 10^7$\,s). By this time the reverse shock has crossed the portion of the jet shown, while the region near the front of the jet is gradually entering a self-similar evolution.}
  \label{fig:jet1D} 
\end{figure}

Figure \ref{fig:jet1D} shows two snapshots of the 1D jet evolution, corresponding to an early time in the jet evolution, shortly before the RS crosses the ejecta injected at constant luminosity ($t \sim 10^{7}$\,s; {\it top panel}), and a later time (\corr{$t \sim 4.3\times 10^{7}$}\,s; {\it bottom panel}).  At early times both the forward and reverse shocks are still close to the head of the jet.  Behind the dense jet head, which is produced during the initial phase of constant jet luminosity, the density decreases to a minimum at $\sim 1.5\times 10^{17}$\,cm.  At yet smaller radii, the density then again increases $\propto r^{-2}$ (eq.~[\ref{eq:rhoj}]) approaching the point of jet injection.  In the more physical 2D case this low-density tail leads to the collapse of the jet channel by the cocoon ($\S\ref{sec:2Dhydro}$).

The relativistic hydrodynamical equations are invariant to an overall change in the normalization of the density and length scale \corr{\citep{Scheck:2002MNRAS}}. Equation (\ref{eq:rhoj}) shows that an increase in the jet luminosity $L_{\rm j}$ (and hence $E_{\rm iso}$) is equivalent to an increase in the normalization of the CMN density $n_{18}$ \corr{for a fixed value of $\eta_R$}.  Because changing $E_{\rm iso}/n_{18}$ {\it does} make a difference in the calculated radio emission, we exploit this scaling freedom to perform a limited scan of the parameter space without having to perform additional simulations.  

\begin{figure}
  \centering
  \includegraphics[width=8.4cm]{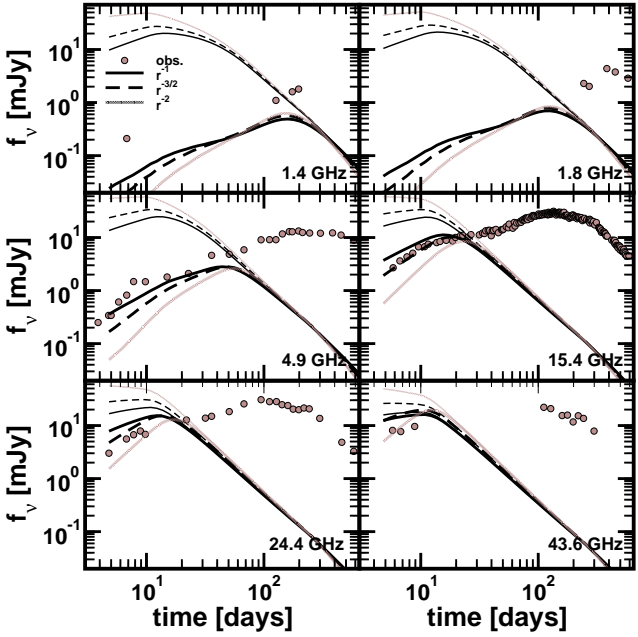}
  \caption{Radio light curves calculated for our fiducial ``fast, narrow" jet model ($\S\ref{sec:FNjet}$), with $E_{\rm iso} = 3.33\times 10^{53}$\,erg, $n_{\rm 18} = 5$ cm$^{-3}$, $\epsilon_e = 0.2$, $\epsilon_B = 2\times 10^{-3}$. \corr{Thick} and \corr{thin} lines show the light curves with and without synchrotron self-absorption respectively.  Different panels showing different observing frequencies, \corr{with} line \corr{styles} showing different assumptions about the power-law slope of the radial density profile of the CNM, $n_{\rm cnm}(r) \propto r^{-k}$, where $k = 1$ (\corr{\it full lines}), 3/2 (\corr{\it dashed}), 1 (\corr{\it grey}).  Shown for comparison with circles are the radio observations of \ubs~(\figref{fig:radio}).}
  \label{fig:FNjet} 
\end{figure}

Figure \ref{fig:FNjet} shows our results for the synthetic radio light curves in comparison to the observations of \ubs~(\figref{fig:radio}). We set $n_{18} = 5$ cm$^{-3}$, which sets $E_{\rm ISO} = 3.33\times 10^{53}$ erg.  Each panel corresponds to a different radio frequency, with different values of the assumed CNM density index $k$ shown with different colours. {\corr{Thick} and \corr{thin} lines show light curves with and without synchrotron self-absorption taken into account}. As expected, the early-time observations can be fit reasonably well using the $r^{-3/2}$ profile (MGM12), but the late-time emission is clearly underproduced. The radio emission is strongly absorbed at $\simless 20$ GHz during the first $10-15$ days, and the peak at each frequency corresponds to the transition from optically thick to optically thin regime. Furthermore, at each frequency we can see the achromatic break in the optically thin light curves (\corr{thin} lines).  As discussed in $\S$~\ref{sec:obs}, this break marks the jet deceleration time.

\begin{figure}
  \centering
  \includegraphics[width=8.4cm]{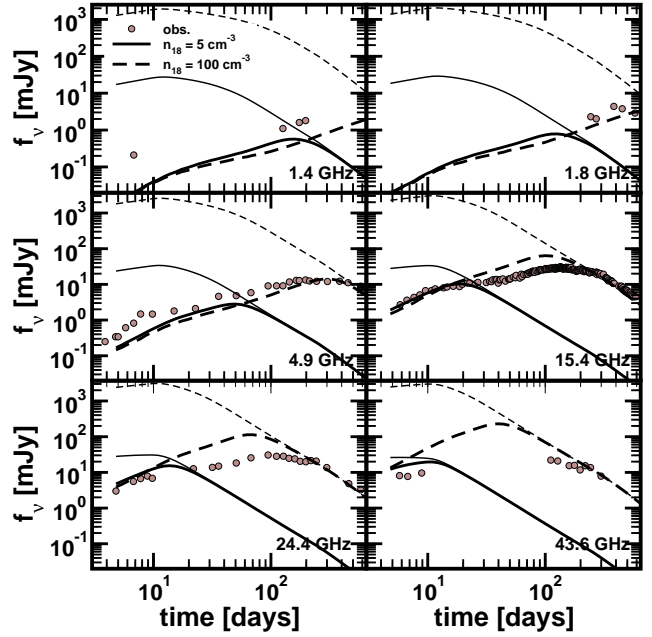}
  \caption{Same as Figure \ref{fig:FNjet}, comparing radio light curves from a jet with higher density and energy ($E_{\rm iso} = 6.67 \times 10^{54}$ erg; \corr{\it dashed line}) that better fits the late-time radio data to the $E_{\rm iso} = 3.33\times 10^{53}$ erg model from Figure \ref{fig:FNjet} (\corr{\it full line}) that better fits the early data.  In both cases we assume an external density profile $n_{\rm cnm} \propto r^{-3/2}$.}
  \label{fig:late} 
\end{figure}

Figure \ref{fig:late} shows the light curves for a case in which $n_{18}$ is increased to $100$ cm$^{-3}$ (and hence $E_{\rm iso}$ to $6.67\times 10^{54}$ erg).  Although the late radio emission is now reasonably well fit, the early-time emission is somewhat overproduced. However, such narrow jets suffer from latteral jet speading effects at rather early stages in their evolution that are not probed by 1D models.  In Sect. 4.3, we explore a similar, fast/narrow jet model with 2D simulations demonstrating that, as a result of latteral spreading, the jet emission is greatly underpredicts observations on timescales longer than $\sim 1$month. 
We have not been able to find any  1-component jet model capable of explaining the late radio rebrightening.

\subsubsection{X-ray cooling in the fast jet} 
\label{sec:Xraycool}

\citet{Kumar:2013aa} propose that the flat radio light curve in \ubs~results from an early-time deficit of the synchrotron emission caused by IC cooling by X-ray photons from the jet.  We test this idea by running jet simulations including radiative losses in equation (\ref{eq:cool}) according to
\begin{equation}
u'_{\rm rad} = \frac{L_x(t_{\rm obs})\Gamma}{4\pi r^2 c},
\end{equation}
where $\Gamma$ is the bulk Lorentz factor of the emitting particles and $L_x$ is the X-ray luminosity, which scales in the same way as the the jet power (eq.~[\ref{eq:lum}]), i.e. according to
\begin{equation}\label{eq:LX}
L_x(t_{\rm obs}) = L_{x, 0}\max\left(1, \dsfrac{t_{\rm obs}}{t_{\rm j}}\right)^{-5/3},
\end{equation}
where again $t_{\rm j} = 5\times 10^5$\,s and we consider several values for the overall normalization, $L_{\rm x,0} = 0$, $5\times 10^{47}$\,erg\,s$^{-1}$, and $10^{48}$\,erg\,s$^{-1}$.  The observer time $t_{\rm obs}$ is related to the simulation time $t$ used in equation (\ref{eq:lum}) according to
\begin{equation}\label{eq:tobs}
t_{\rm obs} = t - \dsfrac{r\cos\theta}{c}\, ,
\end{equation}
where $r$ and $\theta$ are the radius and the angle with respect to the line of sight of the emitting particle at time $t$. 

\begin{figure}
  \centering
  \includegraphics[width=8.4cm]{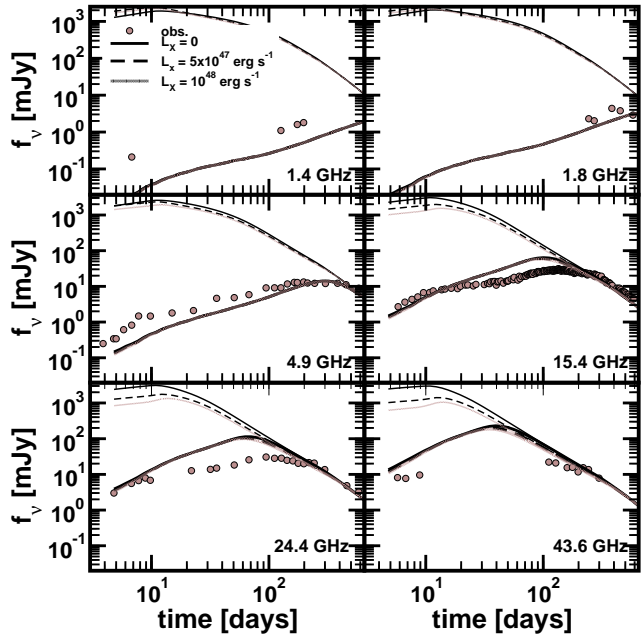}
  \caption{Radio light curves including the effects of X-ray cooling ($\S\ref{sec:Xraycool}$), calculated for a jet with $E_{\rm iso} = 6.67\times 10^{54}$\,erg expanding into a $k=1.5$ medium for different values of the X-ray luminosity $L_{\rm x,0}$ (eq.~[\ref{eq:LX}]).
While X-ray cooling can suppress the synchrotron emission for an optically thin medium (\corr{thin} lines), its effect on the 
actual emission is negligible because the latter is self absorbed (\corr{thick} lines).}
  \label{fig:cool} 
\end{figure}

Figure \ref{fig:cool} shows the radio light curves including IC cooling, calculated assuming a powerful jet ($E_{\rm iso}=6.67 \times 10^{54}$\,erg) and parameters ($k=1.5$, $\epsilon_e = 0.1$, $\epsilon_B = 0.002$) as necessary to match the late radio peak.  IC cooling does noticeably influence the optically thin synchrotron emission.  However, \corr{the effect on the observed, optically-thick emission is almost negligible}, even for the highest assumed value of $L_{x,0}$.  When the emission is strongly absorbed the observer only sees the newly injected particles close to the external shock, while the particles that have cooled are already located further behind and their synchrotron emission does not escape. We thus conclude that IC cooling, within a narrow-jet model, cannot explain the late radio rebrightening.

\subsubsection{Slow, wide jet}
\label{sec:SWjet}

\begin{figure}
  \centering
  \includegraphics[width=8.4cm]{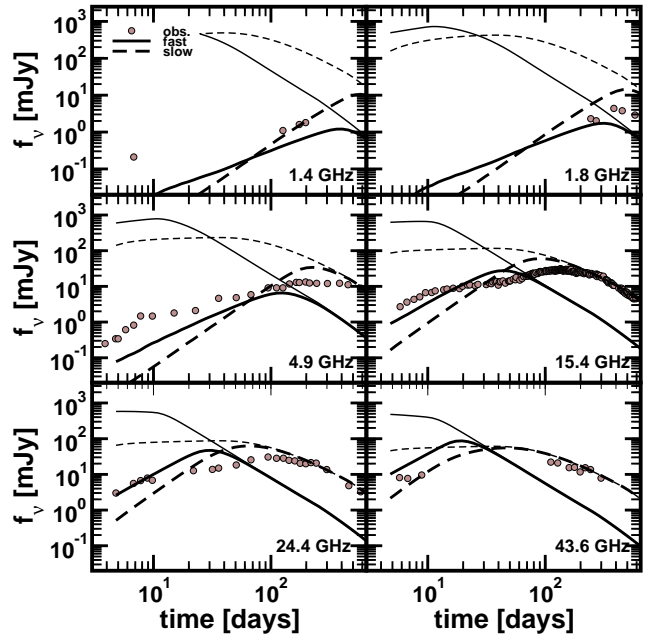}
  \caption{Comparing radio light curves from a slow-wide jet ($\Gamma_{j}=2,\ \theta_{j}=0.5$; {\it \corr{dashed} lines}) to the fast-narrow jet ($\Gamma_{j}=10,\ \theta_{j}=0.1$; {\it \corr{full} lines}).  \corr{Thick and thin} lines show models including and neglecting self-absorption, respectively. Both jets have $E_{\rm iso}=3.33\times 10^{54}$ erg and propagate into a $n_{\rm cnm} \sim r^{-3/2}$ external medium with $n_{18} = 50$ cm$^{-3}$. We assume $\epsilon_e = 0.1$ and $\epsilon_B = 0.001$.}
\label{fig:SWjet}
\end{figure}

To contrast with the fast, narrow jet model (Sec.~\ref{sec:FNjet}) we also consider a slow, wide jet with $\Gamma_{j}=2$ and $\theta_{j}=0.5$ (Sec.~\ref{sec:FNjet}).  Figure \ref{fig:SWjet} shows our results for the slow jet light curves ({\it \corr{dashed} lines}), calculated for an external density $n_{18}=50$ cm$^{-3}$ and jet energy $E_{\rm iso} = 3.33 \times 10^{54}$ erg, in comparison to the equivalent fast jet case ({\it \corr{full} lines}).  The slow jet can explain the late time peak, but it underpredicts the early-time emission observed from \ubs.  This is because a slow jet coasts for a long time before appreciably slowing down, since it must first sweep up $\sim 1/\Gamma_{\rm j}$ of its own rest mass.  During this extended coasting phase, the emitting area increases and so does the luminosity of the source at bands where the emission is self-absorbed. The resulting lightcurve rises fast before reaching its peak, with the peak time determined by either the thick-thin transition or the onset of the jet deceleration, by itself inconsistent with the relatively flat light curve observed for \ubs.  \corr{The comparison between the light curves produced by a fast-narrow jet and a slow-wide jet (Fig.~\ref{fig:SWjet}) is suggestive of the fact that different parts of a unique jet, moving at different angles and with different Lorentz factors, could account for both the early and the late time light curve of \ubs.}

\subsection{Two-component 1D models}
\label{sec:1D2Z}

The one-component jet models considered thus far can reproduce either the early or late time light curves well \corr{(\figref{fig:SWjet})}, but none acceptably reproduce the \emph{entire} emission.   We are thus motivated to consider a two-component jet, comprised of a fast core ($\Gamma_{\rm j} = 10$) with a narrow half-opening angle of $\theta_{\rm j} = 0.1$ rad, which is surrounded by a slower sheath ($\Gamma = 2$) from $0.1$ to $0.5$ rad.  Both components of the jet are assumed to possess an equal isotropic energy $E_{\rm iso} = 4\times 10^{54}$ erg.  We employ an external medium with a $n_{\rm cnm} \propto r^{-1.5}$ profile and $n_{18} = 60$ cm$^{-3}$.  As in previous cases we assume $\epsilon_e=0.1$, however we allow $\epsilon_B$ to vary modestly about its fiducial value $10^{-3}$ so as to quantitatively minimize the differences among observational data and our synthetic models.

\begin{figure}
  \centering

  \includegraphics[width=8.4cm]{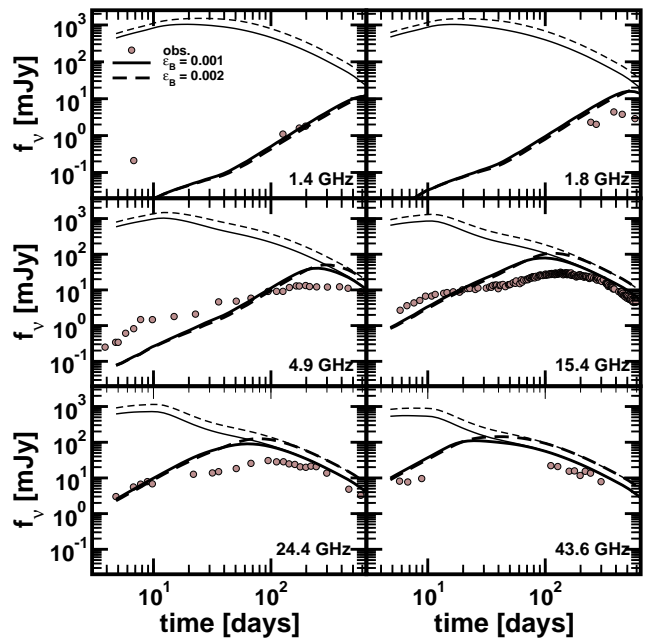}
  \caption{Light curves of a two-component 1D jet model ($\S\ref{sec:1D2Z}$) including a fast core ($\Gamma_{\rm j} = 10$; $\theta_{\rm j} = 0.1$) surrounded by a slower sheath ($\Gamma_{\rm j} = 2$; $\theta_{\rm j} = 0.5$, each with the same isotropic energy $E_{\rm iso} = 4\times 10^{54}$ erg.  Different \corr{line styles} show different choices of the assumed post-shock magnetic field strength $\epsilon_B = 10^{-3}$ ({\it \corr{full} line}) and $2\times 10^{-3}$ ({\it \corr{dashed} line}).}
  \label{fig:twocomp} 
\end{figure}

Figure \ref{fig:twocomp} shows the results of our 2-component 1D model.  The model with $\epsilon_B=10^{-3}$ provides a reasonable representation of the well-sampled $\nu\simmore 15$\,GHz light curve, but it underpredicts the flux at low frequencies $\nu\lesssim 5$\,GHz and early  times $t\lesssim 10$ days.  This excess at early times \corr{in the observational data} appears to require \corr{either a significant reduction of the optical depth at  times $t\lesssim 10$ days}, \corr{or an additional emitting region not included in the calculations, and quite likely, a combination of both factors. \cite{Mimica:2009aa} showed that during the early stages of the jet-CNM interaction the reverse shock that decelerates the jet material can be particularly powerful. Indeed, for the adopted parameters, the  rate of dissipation of energy at the reverse shock peaks on a timescale comparable to the observed jet duration ($\sim 10$ days in the case of \ubs). However, for the fiducial parameters adopted in the present paper (including those of the CNM), the emission of the reverse shock is strongly absorbed below $\sim 10$\,GHz and, hence, the reverse shock contribution alone cannot account for the flux deficit of our models at early times\footnote{\corr{Besides the fact that it is technically very challenging to accurately detect the reverse shock in the hydrodynamic simulations, the reverse shock emission is strongly suppressed for the parameters at hand, which justifies our choice of not including it in the current models.}}. The lack of a dominant reverse shock component at early times is potentially supported by the upper limits of $\lesssim 2-10\%$ on the 4.8 GHz linear polarization on timescales of weeks after the BAT trigger (\citealt{Wiersema+12}).  We note, however, that if we consider a smaller value of the $\zeta_e$ parameter, the synchrotron self-absorption at both the forward and reverse shocks is substantially reduced, and even counting only the forward shock emission, the flux deficit at early times can be ameliorated (see Appendix~\ref{app:zeta_e} for details).}  

The model with $\epsilon_B = 0.002$, though slightly overpredicting the observed emission, nevertheless provides a good candidate to set the input for our full 2D simulations described next.  It is desirable to slightly overpredict the emission with the 1D model since in the full 2D case the jet will experience lateral spreading.  This leads to earlier deceleration of the jet in comparison to spherical models, resulting in lower flux at bands where the emission is optically thin (see next Section).

\subsection{2D simulations}\label{sec:2D}

Motivated by the improvements in accommodating the light curves when considering one-dimensional two component models (\figref{fig:twocomp}), we perform 2D simulations \corr{of two component jet models} with $E_{\rm iso} = 4\times 10^{54}$ erg. The jet is injected at the inner grid boundary ($r_{j, 0} = 2\times 10^{16}$ cm) in the angular range $[0, 0.5]$ rad. A fast core ($\Gamma_f = 10$) is injected between $0$ and $0.1$ rad, while a slower sheath ($\Gamma_s = 2$) is injected between $0.1$ and $0.5$ rad. The external medium into which the jet is propagating has a $r^{-3/2}$ profile and $n_{18} = 60$ cm$^{-3}$. The jet luminosity changes in time according to Eq.~\ref{eq:lum}.

The radial resolution of the simulations is $\Delta r = 4\times 10^{14}$ cm (Sec.~\ref{sec:2Ddetails}), but the radial grid is regularly extended as the jet propagates into the external medium. This is achieved by stopping the simulation once the jet head nears the outer radial edge of the current grid, extending the radial grid and initialising with the external medium. The angular resolution is $\Delta \theta = \pi/600$ rad and we always simulate the full range $[0, \pi/2]$ using $300$ points.

In order to compute the light curves the snapshots of the grid state need to be saved periodically (Sec.~\ref{sec:1Ddetails}). The time at which the snapshots are written is exactly the same as in 1D simulations, thus enabling us to directly compare 1D and 2D models both in terms of their hydrodynamical evolution (Sec.~\ref{sec:2Dhydro}), as well as in their observational signatures (Sec.~\ref{sec:2Dlcs}).

\subsubsection{Hydrodynamic evolution}
\label{sec:2Dhydro}

\begin{figure}
  \centering
  \includegraphics[width=8.6cm]{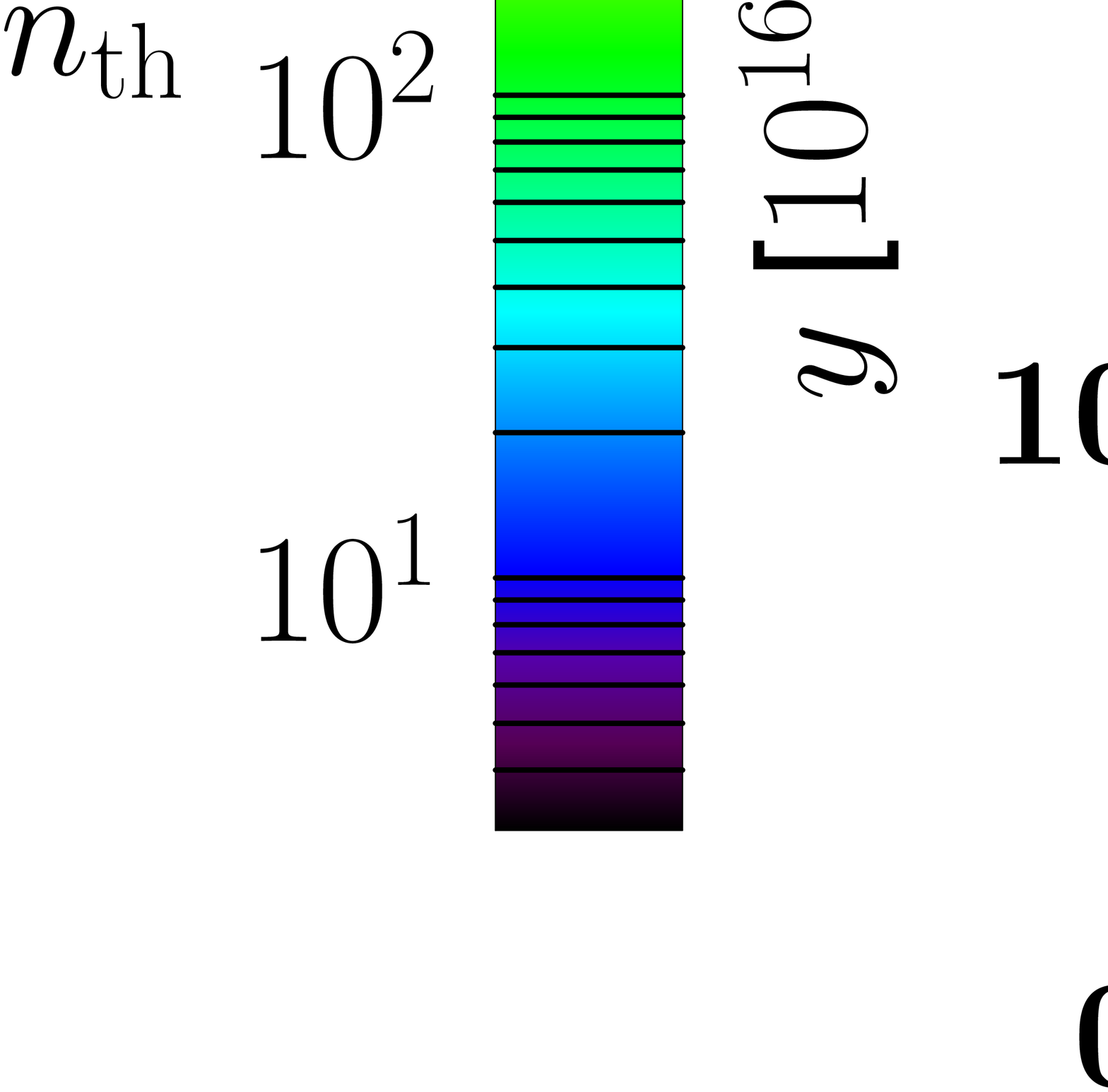}
  \includegraphics[width=8.6cm]{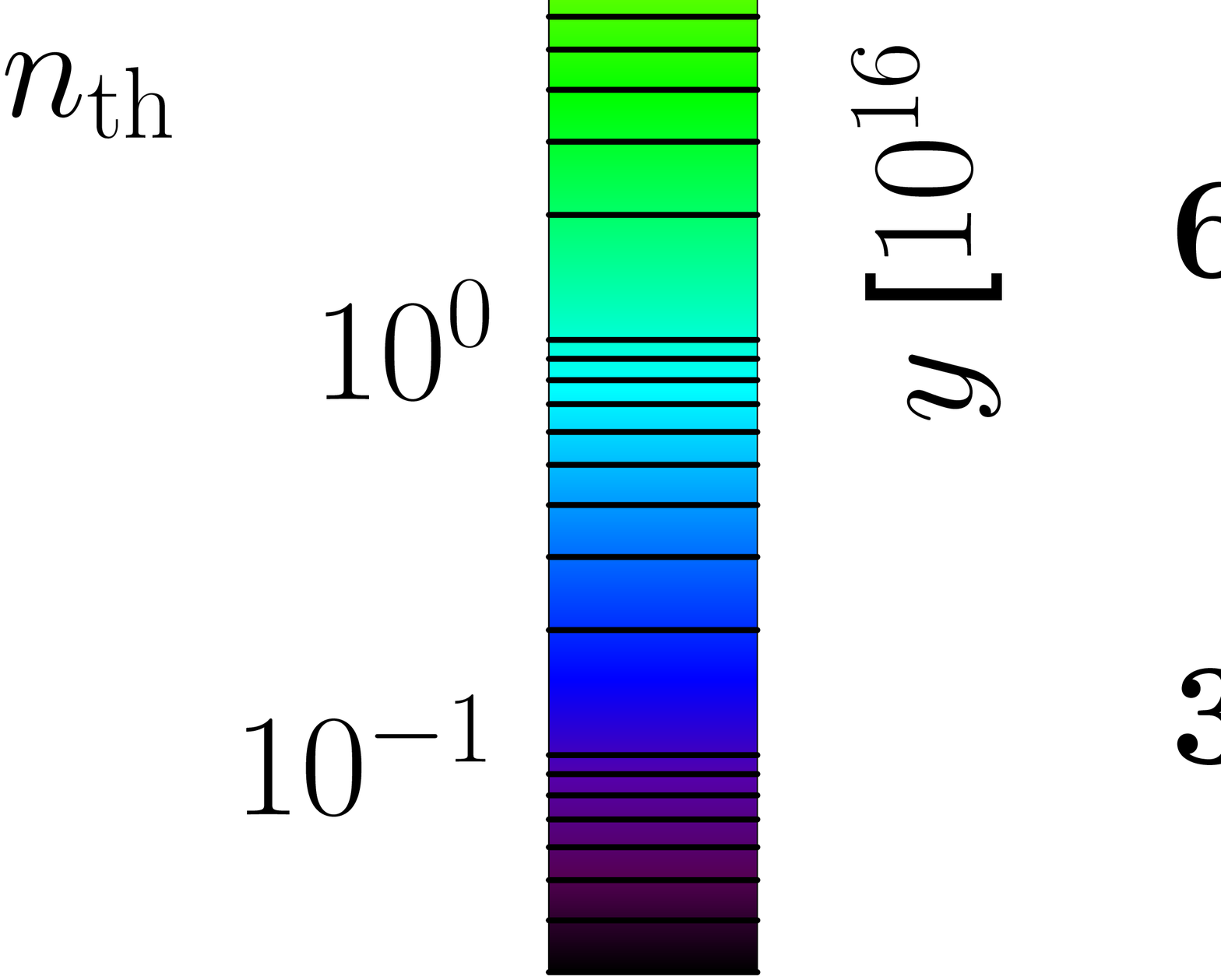}
  \includegraphics[width=8.6cm]{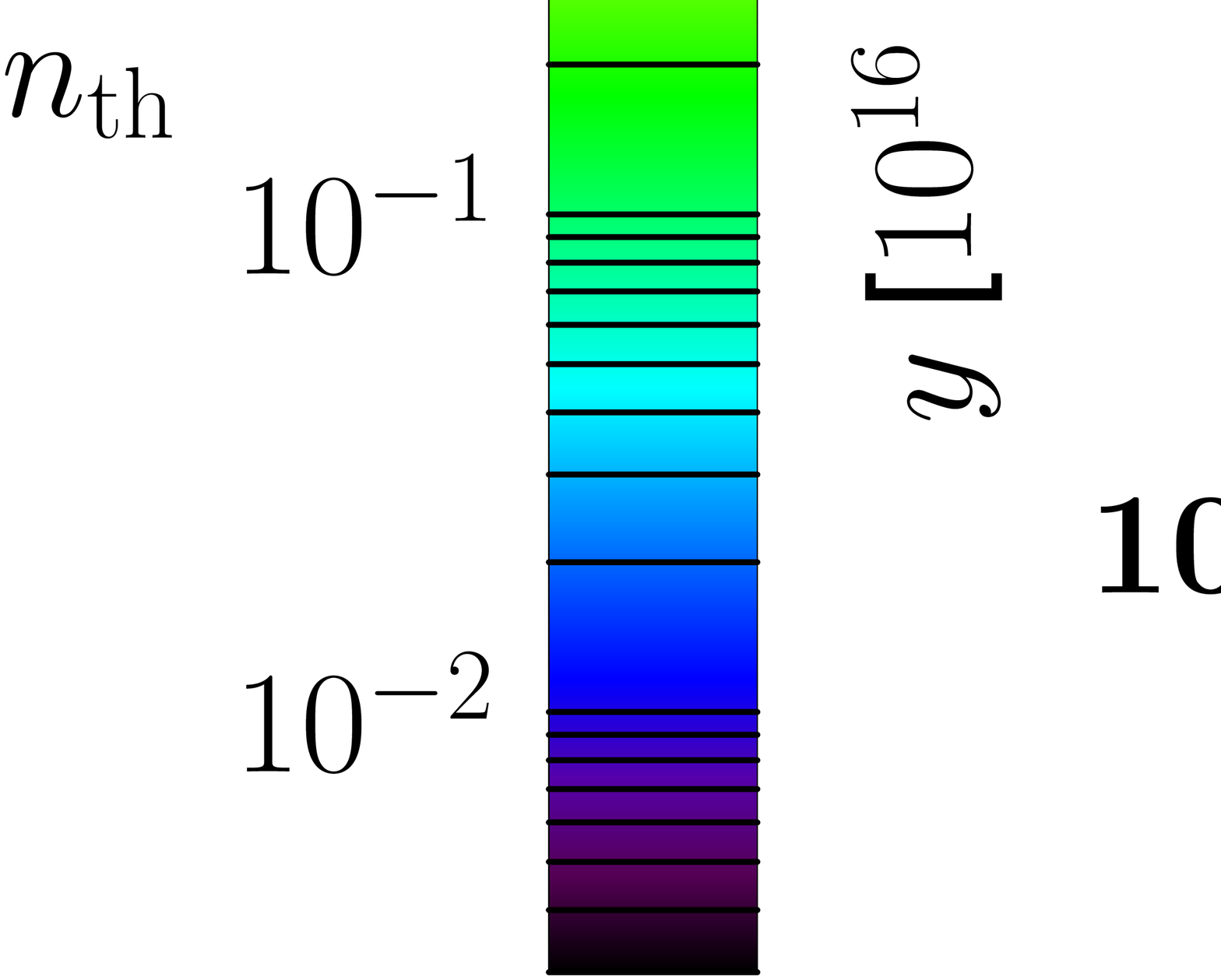}
  \includegraphics[width=8.6cm]{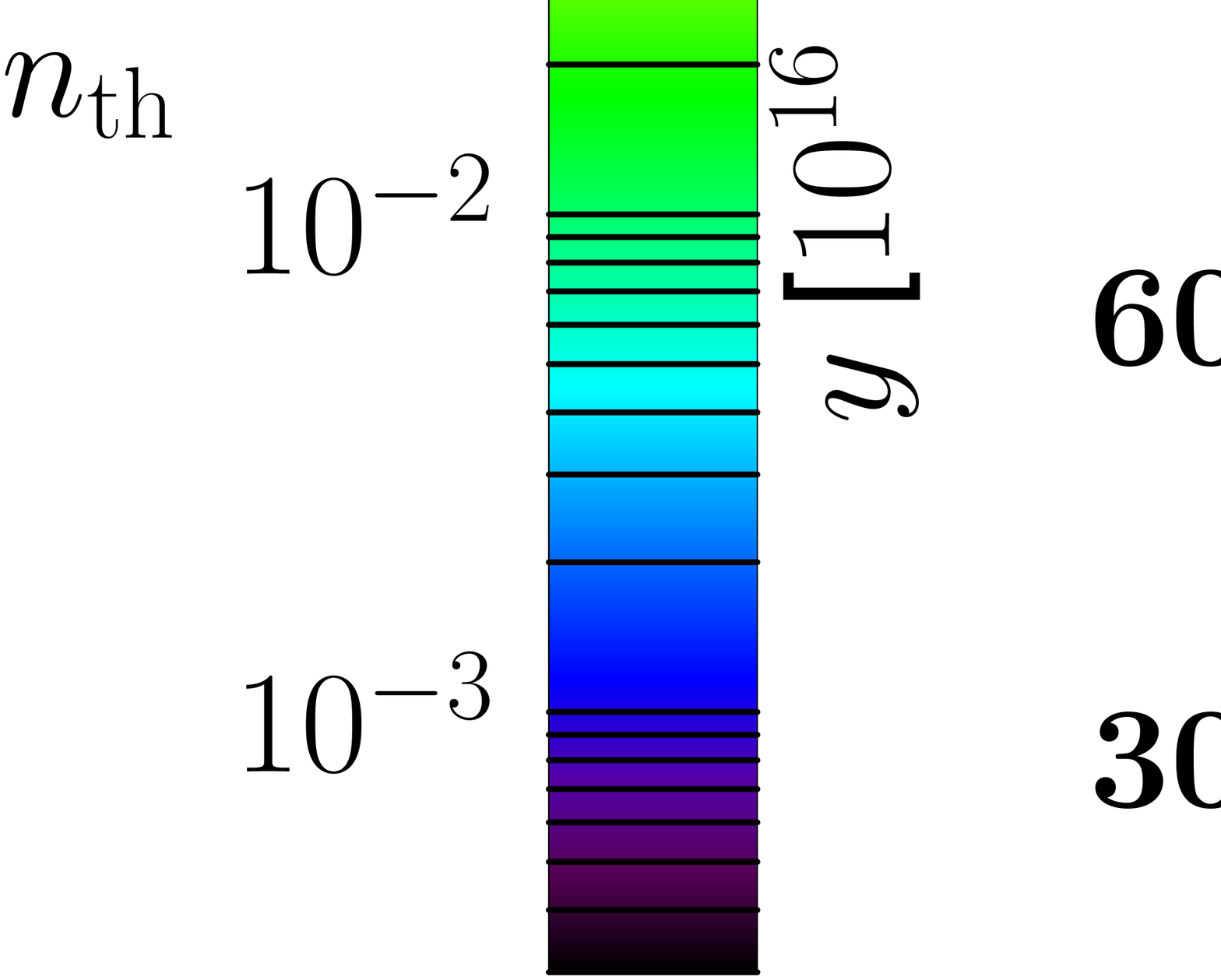}
  \caption{Snapshots of the hydrodynamic evolution of the two-component jet. The time in the upper two panels corresponds to the times of the snapshots in \figref{fig:jet1D} and \figref{fig:jet1D2D}. In each panel the left half shows the (comoving) number density $n_{\rm th}$ of the thermal fluid, while the right half shows the fluid four-velocity $\Gamma\beta:=\Gamma v/c$. Since both fast ($\Gamma_f = 10$) and slow ($\Gamma_s = 2$) jets are injected with the same luminosity per solid angle, $n_{\rm th}$ in the (more relativistic) fast jet is initially $\sim \Gamma_s^2 / \Gamma_f^2$ lower than in the slower jet. This can be seen in the upper two panels, where the fast jet is seen as a dark-coloured channel close to the symmetry axis. Note that the density color scale changes from panel to panel. The spherical boundaries on the left side of some panels denote the edge of a radial grid which is periodically extended, see \secref{sec:model}.}
  \label{fig:jet2D} 
\end{figure}

\begin{figure}
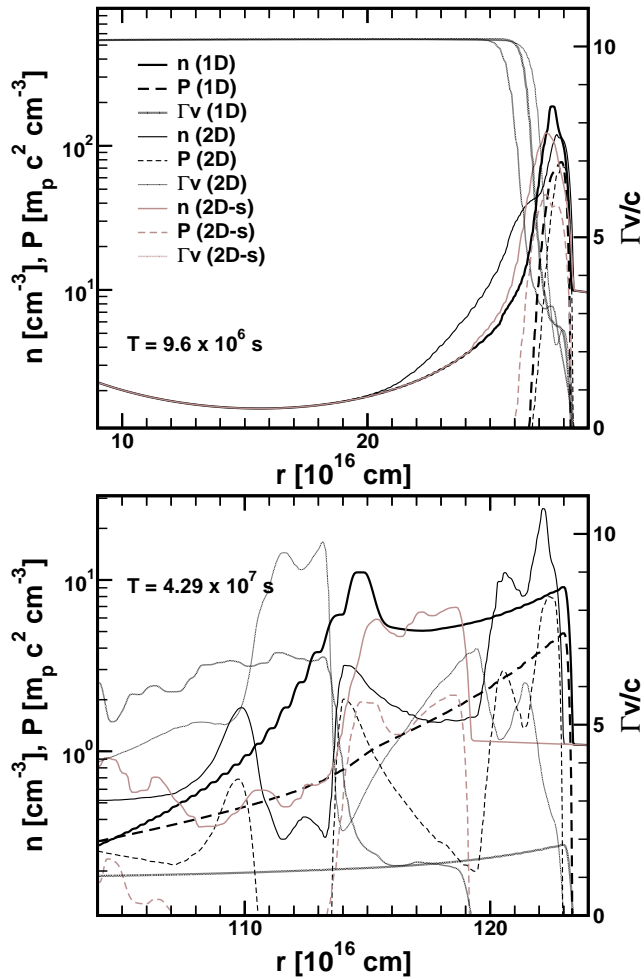

  \centering
  \includegraphics[width=8.4cm]{figures/hydro1D2D-early.eps}
  \includegraphics[width=8.4cm]{figures/hydro1D2D-late.eps}
  \caption{Same as \figref{fig:jet1D}, but in addition showing the values at the axis of the two-component 2D jet simulation (\corr{thin} lines), and a 2D simulation of a single-component fast jet (\corr{brown} lines). For better comparison, results of a 1D simulation with the same $\Delta r$ as the 2D one are shown (see Sec.~\ref{sec:2Ddetails}).}
  \label{fig:jet1D2D} 
\end{figure}

Our simulation starts from sufficiently short distance from the central engine so that we can follow all the stages of the jet deceleration as the jet sweeps through the CNM. Initially both the slow and fast jets are overdense with respect to external medium and propagate almost ballistically along their initial opening angle (top panel in \figref{fig:jet2D}). At this stage, both fast and slow jets move radially and are slowed down at the reverse shock (located at distance $\sim 10^{18}$\,cm at the second panel in \figref{fig:jet2D}). The reverse shock strongly compresses and heats the jet material. The shock front can be clearly seen in the right panels of \figref{fig:jet2D} where the unshocked CNM is at rest (black area).  Throughout this work, we assume that particle accelerated at the external shock are responsible for the radio emission of \ubs.

There is a fundamental feature of the structured jet dynamics that deserves special attention: the jet is remarkably stable throughout most of the simulated evolution. An analytical study of the stability of our configuration is rather complex (and beyond the scope of this paper), the reason being that the fast jet is underpressured with respect to the slow one. This breaks out a common assumption in the study of relativistic jets, namely that the jet is pressure-matched with the surrounding matter. As a result of this, the slow / fast jet interface develops a Riemann structure consisting on a pair of shocks separated by a contact discontinuity. For the conditions we study, both shocks move towards the jet axis, though the one shocking the fast jet is faster (and weaker) than the one crossing the beam of the slow jet. Across the faster shock, the jet Lorentz factor slightly decreases from $\Gamma_f=10$ to $\Gamma\sim 9$, while across the slower shock it decreases almost to the slow jet value ($\Gamma_s=2$). In order to estimate the shock front velocities, it is important to consider that the fast jet is launched to expand radially at an angle $\theta_j=0.1$ and, thus, with a velocity component perpendicular to the jet axis $v_\perp \sim \theta_j (1 - \Gamma_f^{-2})^{1/2} = 0.1 c$. Balancing the fast and slow jet ram pressure perpendicular to the axis we conclude that the shock propagation speed towards the axis is very slow, while the effect of the initial (thermal) pressure imbalance between both jets gets reduced to a relatively thin layer around the fast jet (see, e.g., the narrow red layer surrounding the beam of the fast jet -white shades- in the right side of the second panel of  \figref{fig:jet1D2D}).

Having explained the differences with previous jet stability studies caused by the pressure imbalance, we now use the results of those works to analyze fast/slow jet interface properties. \cite{Hardee:2003aa} showed that pressure matched, cylindrical jets with a fast-spine and a slow-sheath are more stable than jets without slow sheaths.  In our 2D model the role of the sheath can be played by the slow jet. \cite{Hardee:2003aa} show that the decrease in the growth rate of Kelvin-Helmholtz instabilities is proportional to the velocity shear $\Delta v$. In our case, $\Delta v = v_{\rm fj} - v_{\rm sj}\simeq 0.13c$ ($v_{\rm fj}$ and $v_{\rm sj}$ are the speeds of injection of the fast and of the slow jets, respectively). They also find that the presence of an external moving shear dampens asymmetric modes and raises the growth rate of the fundamental pinch mode. In this regard, we note that, initially ($T\lesssim 100\,$days) the fast jet does not develop any recollimation induced by the pinching of the slow jet. Later ($T\gtrsim 100\,$days), the bow shock and the reverse shock produced by the slow jet pinch the beam of the fast jet only very close to the jet head ($y \sim 2.8\times 10^{17}\,$cm; top panel of \figref{fig:jet2D}). This pinching speeds up the beam of the fast jet which, as a result, propagates a bit faster than a pure 1D ``jet'' with the same properties. This effect is visible in the top panel of \figref{fig:jet1D2D}, where profiles along the axis of the 2D jet model are compared with an equivalent 1D jet model. There we observe that the forward shock of the 1D jet model lags behind that of the 2D jet (located at $r\simeq 2.9\times10^{17}\,$cm). 
Also in \figref{fig:jet1D2D}, we display the axial profiles of a 2D jet with the same properties as the fast and narrow jet, but without a broader slow jet flanking it (``top-hat'' 2D jet). Without the stabilising influence on the dynamics of the broader, slower jet, the top-hat 2D jet clearly lags behind 1D jet model. At later times (bottom panel of \figref{fig:jet1D2D}), the situation is reversed, and the 2D effects (transfer of transversal momentum) makes the two-component jet bow shock to propagate at smaller speed. 

At this point the ram pressure of the fast jet is $\sim 3$ times larger than that of the slow jet and the former ``breaks-out'' of the latter's bow shock, as can be observed from the third panel of  \figref{fig:jet2D}. In the subsequent evolution, the initially oblate structure of the outermost bow shock of the fast jet becomes increasingly prolate. After $T \sim 20\,$yr (bottom panel of \figref{fig:jet2D}), the longitudinal sizes of the cavity blown by the fast jet and the slow jet cavity become comparable.
 
Another important difference we find between a top-hat 2D jet model and our two-component jet model is the radial width of the bow-shock-to-reverse shock region. As time evolves this radial width becomes shorter in the top-hat jet, since there is no dynamical collimation exerted by the broad component. For instance, at $T\sim 500\,$days (bottom panel of \figref{fig:jet1D2D} and second panel of \figref{fig:jet2D}), the radial width of the former region is of $\sim 5\times 10^{16}\,$cm for the top-hat 2D jet model, while it is about twice as large in our reference sheared 2D jet. This difference increases with time, particularly after the constant injection phases finishes.

At later times the injection luminosity in both jets decreases. The tenuous jet finds itself surrounded by a hot cocoon of shocked material. The pressure of the surrounding cocoon causes the jet channel to collapse (third panel in \figref{fig:jet2D}). At later times very little new injection of energy and momentum takes place.  The blast wave that bounds the structure of the complex cavity will relax into a spherical Sedov solution. While we do not follow its evolution until this late stage, the resulting emission (of interest here) is practically spherically symmetric by the end of the simulation (see next Section).

\subsubsection{Light curves}
\label{sec:2Dlcs}

\begin{figure}
  \centering
  \includegraphics[width=8.4cm]{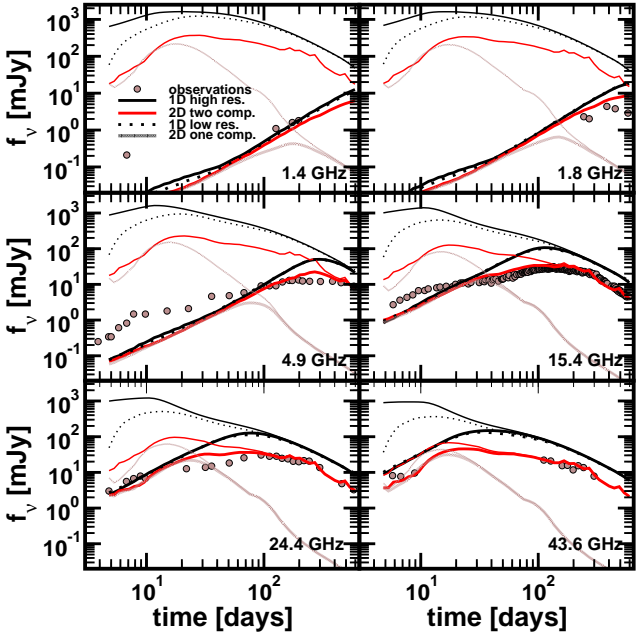}
  \caption{Same as \figref{fig:twocomp}, but showing the light curves produced by 1D simulations (\corr{full} lines, corresponding to \corr{dashed} lines in \figref{fig:twocomp}) and 2D simulations of a two-component jet (\corr{red} lines) for $\epsilon_e=0.1$, $\epsilon_B = 0.002$, \corr{as well as a 2D simulation of one-component fast and narrow jet (grey lines).} Both 1D and 2D simulations assume $E_{\rm iso} = 4\times 10^{54}$ erg, an external medium with $r^{-3/2}$ profile and $n_{18} = 60$ cm$^{-3}$. The fast and narrow jet ($\Gamma_f = 10$) spans the angle from $0$ to $0.1$ rad, while the slow and wide jet ($\Gamma_s = 2$) extends from $0.1$ to $0.5$ rad.}
  \label{fig:onetwo} 
\end{figure}

Figure~\ref{fig:onetwo} shows a comparison between light curves computed from 1D and 2D simulations with the same initial conditions, \corr{as well as a light curve computed from a 2D simulation of an one-component jet (fast and narrow component only)}.  Since the 2D simulations are performed with lower resolution in the radial direction in comparison to the 1D ones, we have also performed 1D simulations with reduced resolution (to match the 2D radial grid, dotted line). Though, from a dynamical point of view the differences in changing resolution are negligible, the initial speed of propagation of the forward (bow) shock of the jet is slightly different ($\sim 10^{-3}c$; see \citealt{Mimica:2009aa}), which slightly affects the flux arrival times and the optical thickness of the overall ejecta. One can see that the  reduced resolution affects the predicted optically thin emission at early times (\corr{dotted} and \corr{full thin} curves in \figref{fig:onetwo}). However, the emission at these stages is optically thick and the low resolution has negligible effect on the emission (see \corr{thick} lines). We, therefore, expect that the resolution of the 2D simulations is sufficient to capture the hydrodynamic evolution of the jet well enough to compute the resulting emission (see Appendix A for more details on other numerical parameters influencing the emission.)

At low frequencies the emission is absorbed and 1D and 2D jets have very similar light curves as long as the jet is expanding conically. In the 2D \corr{two-component} simulations the emission flattens earlier as a result of the jet deceleration and expansion along the azimuthal direction. In the 1D simulations such expansion is not allowed and the jet deceleration takes place more slowly.
 
At higher frequencies the 2D \corr{two-component} jet emission peaks earlier and at lower luminosity. The 1D model overpredicts the optically thin emission by factor of $\sim 3-5$, taking into account that our 1D treatment already takes into account that the opening angle of the jet is finite when calculating the emission. Models that assume 1D, spherically emitting blast overpredict the  resulting emission by a large factor ($\sim 100$) for timescales $\sim$ months after the TDE.
   
\corr{At early times, when all models radiate in an optically thick regime, the differences among 1D and 2D models are negligible, and all of them fall at roughly the same distance to the observations. However, the one-component 2D jet fails badly to account for the observed flux after its light curve peak, when the flux decreases much faster than either the two component 2D model or than the 1D single-component jet models shown in Fig.~\ref{fig:FNjet}. For brevity, we do not show here the light curves obtained with a single component 2D slow-wide jet, but the effects on the light curve are very similar to the ones shown in Fig.~\ref{fig:SWjet} and discussed in Sect.~\ref{sec:SWjet} for 1D models. Considering that a 2D hydrodynamical modeling is more realistic than any 1D modeling, we find that the orders of magnitude flux deficit in single component 2D fast-narrow jet models at late times, together with the flux deficit of single component 2D slow-wide jet modes at early times is an unequivocal proof that a single component jet cannot simultaneously explain both early- and late-time emission, and argues in favor of a two-layer jet model.}

\subsubsection{Origin of non-thermal emission}
\label{fig:nth2D}

As remarked in \secref{sec:nonthermal}, \verb'SPEV' models non-thermal particle evolution by injecting them behind relativistic shock fronts. {Figure~\ref{fig:nth2D} shows the particle distribution and density for the four hydrodynamic snapshots discussed in \figref{fig:jet2D}.

\begin{figure}
  \centering
  \includegraphics[width=8.6cm]{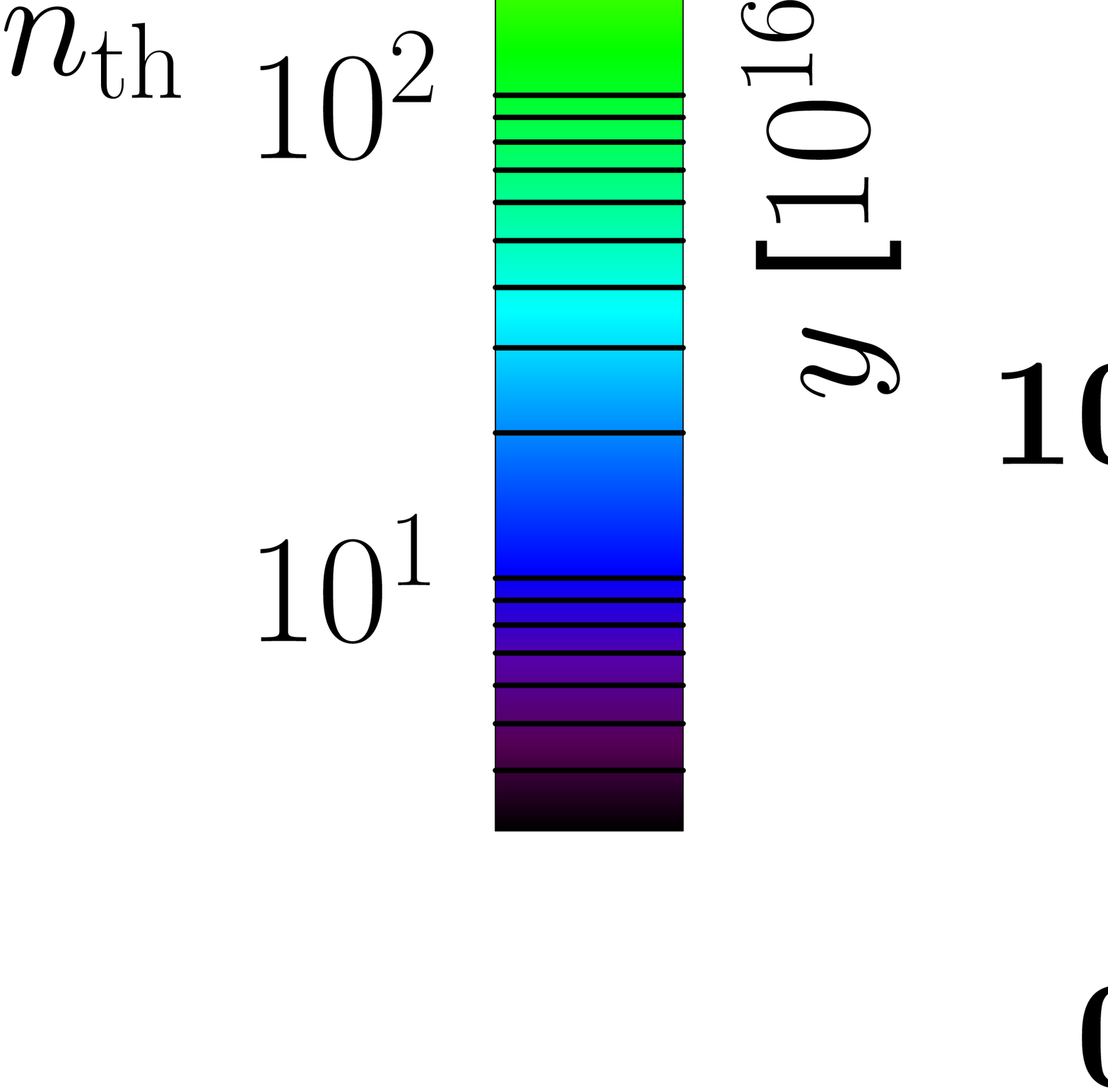}
  \includegraphics[width=8.6cm]{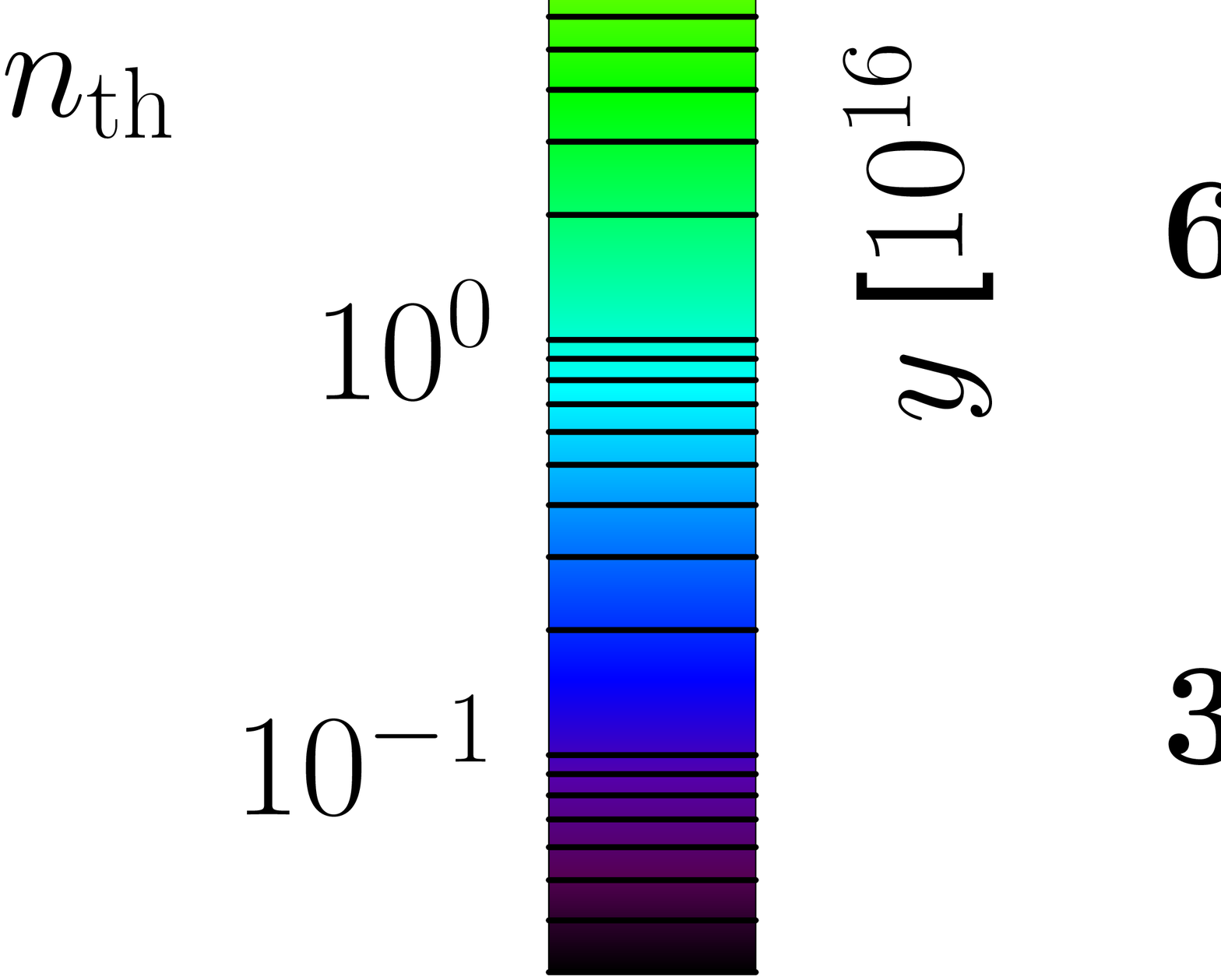}
  \includegraphics[width=8.6cm]{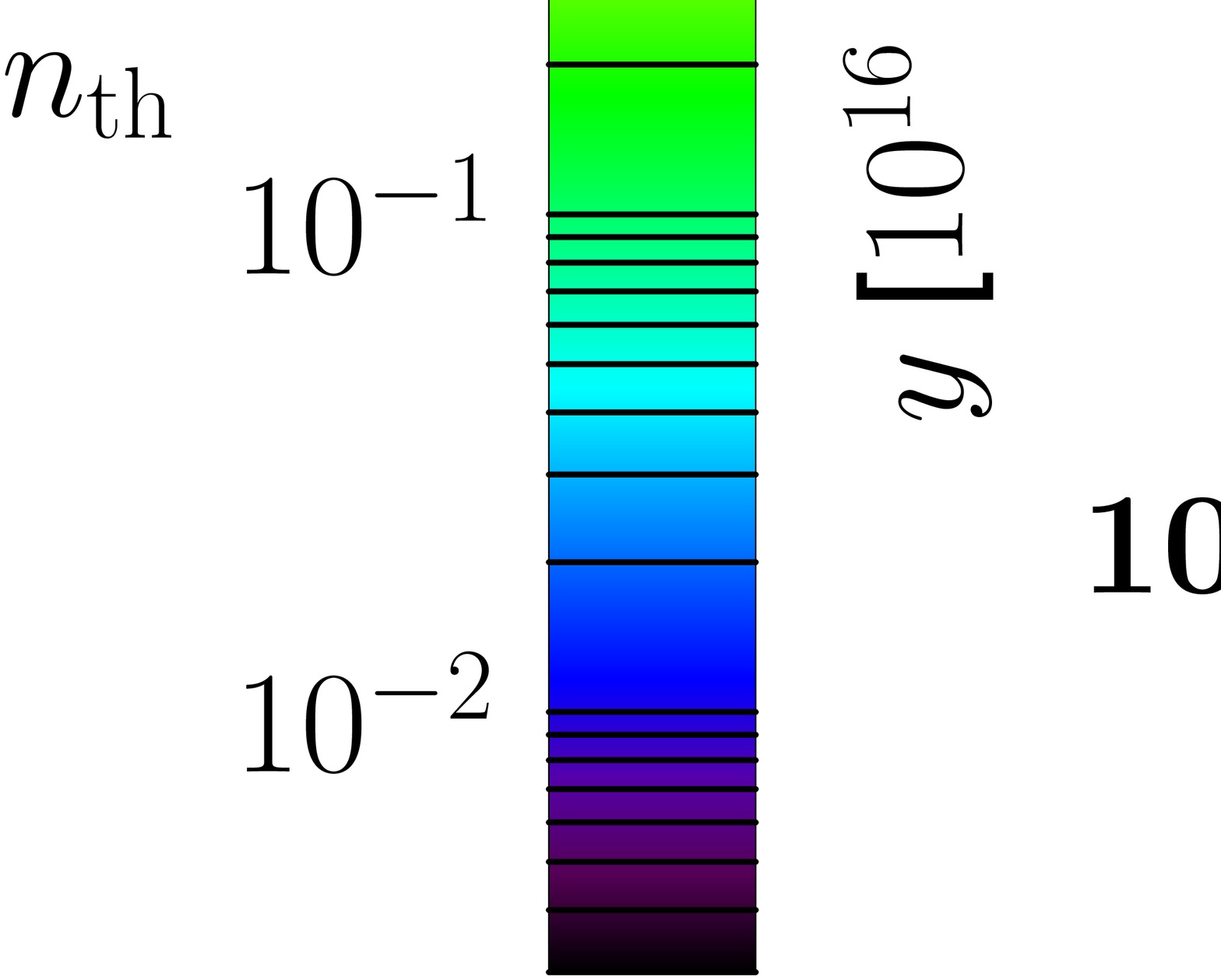}
  \includegraphics[width=8.6cm]{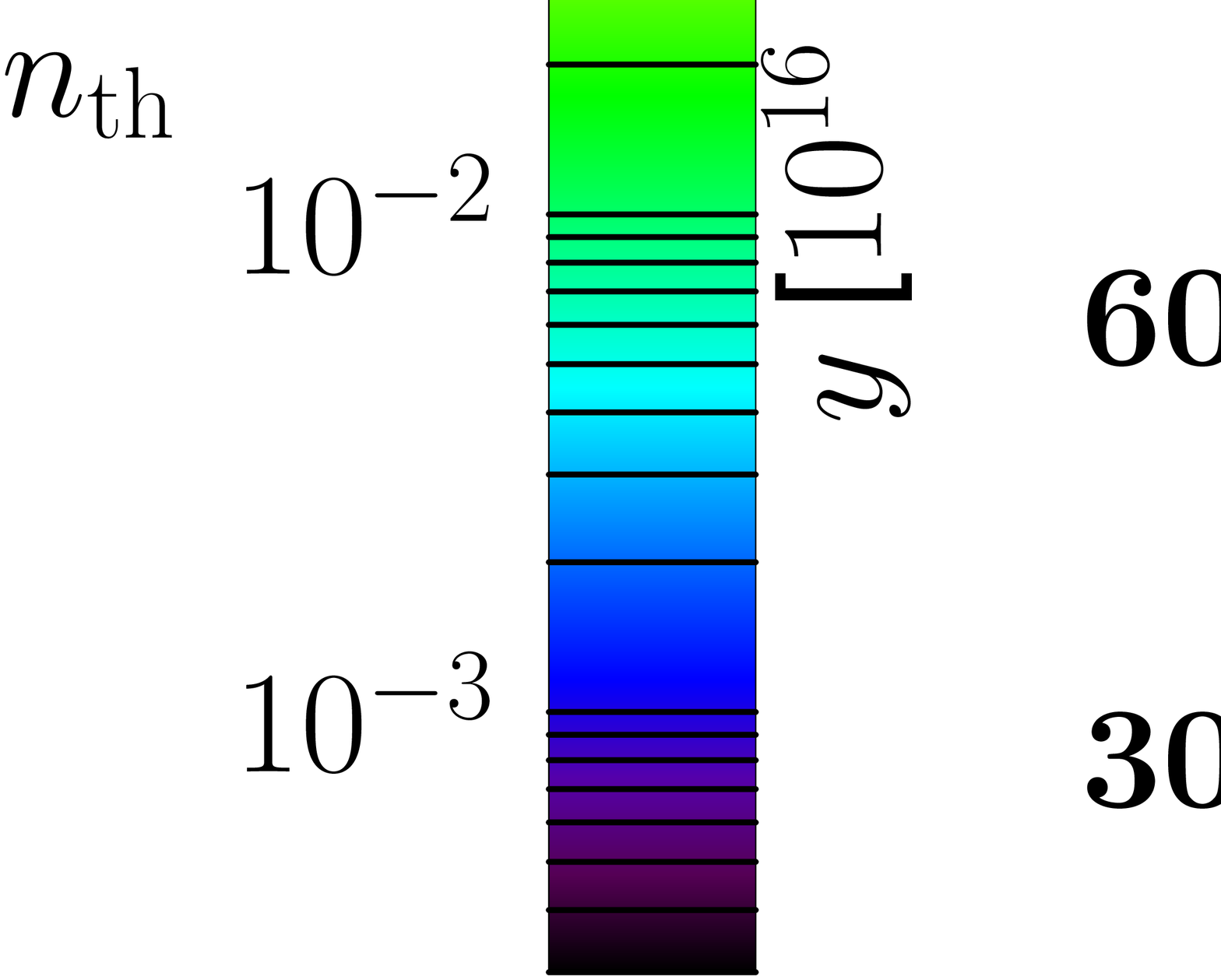}
  \caption{Same as \figref{fig:jet2D}, but showing the number density of the non-thermal particles instead of normalised fluid velocity.}
  \label{fig:nth2D} 
\end{figure}

Even though there is substantial dissipation at the reverse shocks that can contribute to the emission the first $\sim$10\,days, we only model particles accelerated at forward/external shocks. The reason is \corr{threefold: (1) the emission from the reverse shock is strongly dimmed due to the strong synchrotron self-absorption resulting from the adopted microphysical and dynamical parameters of our models; (2) we focus on the month to year timescale evolution where the reverse shock emission is expected to be less prominent; and (3) computing the reverse shock particle evolution and emission is numerically very demanding.}

At early times (top panel in \figref{fig:nth2D}) the particles are confined to a relatively small volume behind the shock fronts and contribute to the observed emission. At later times (bottom two panels in \figref{fig:nth2D}) a significant fraction of particles stream away from the shock front since the distance between the forward shock and the contact discontinuity separating the shocked jet material from the shocked CNM also grows.  It should remarked that we stop injection when $\gamma_{\rm min}$ of the electrons becomes less than $\sim 2$. We have checked that our results do not depend on the exact value of the cutoff. For $t\simmore 10\,$years the blast wave enters the ``deep Newtonian'' regime, where $\gamma_{\rm min} \sim 1$ for all freshly inject electrons and the particle acceleration spectrum is modified (see \citealt{Sironi:2013aa} for details).

\section{Applications and Implications}
\label{sec:app}

\subsection{Properties of super-Eddington accretion flows}

\ubs~provides a testbed to study relativistic jet formation associated with super-Eddington accretion, as may occur in TDEs as well as other astrophysical systems such as ultra-luminous X-ray sources and gamma-ray bursts (GRBs).  Our models show that the radio afterglow of \ubs~is well explained by a structured jet, comprised of a high-$\Gamma_{\rm j}$ core surrounding by a mildly relativistic ($\Gamma_{\rm j} \sim 2$) sheath containing a comparable, or greater, kinetic energy.  

The geometry that we infer could simply reflect the generic structure of jet/accretion disk systems.  If powered by the Blandford-Znajek process, the fast jet core could contain those magnetic field lines threading the SMBH horizon.  The slower sheath could, by contrast, represent a mildly relativistic outflow from the inner accretion disk.  Powerful disk outflows are expected theoretically (e.g., \citealt{Blandford&Begelman99}), as supported by radiation-hydrodynamical (\citealt{Ohsuga+05}) and magneto-hydrodynamical (\citealt{Jiang+14}) simulations of super-Eddington accretion.  

Outflows from the super-Eddington accretion flow are also expected to reduce the mass accretion rate reaching the SMBH horizon at $R_{\rm in} \sim R_{g}$ by a factor $\sim (R_{\rm out}/R_{\rm in})^{p}$, where $R_{\rm out}$ is the outer edge of the disk and $p \sim 1$.  The fact that our modeling demands the sheath possess an energy of $\sim$ few $10^{53}$ erg, comparable to the entire available rest mass $\sim M_{\odot} c^{2}$ erg of the tidally disrupted star, implies that the accreted matter must have been deposited into the disk close to the SMBH.  This is consistent with the TDE explanation for \ubs~because the low angular momentum of the stellar debris causes it to circularise interior to twice the tidal radius $R_{\rm t} \approx R_{\star}(M_{\bullet}/M_{\star})^{1/3} \sim 50(M_{\bullet}/10^{6}R_{\odot})^{-2/3}R_{\rm g}$, where $R_{\rm g} \equiv GM_{\bullet}/c^{2}$, $M_{\bullet}$ is the black hole mass and the numerical estimate is for a solar type star with mass $M_{\star} = M_{\odot}$ and radius $R_{\star} = R_{\odot}$.    

\citet{Tchekhovskoy:2014aa} alternatively propose that the slow sheath could be created by an initially precessing fast jet (cf.~\citealt{Lei+13}), which then aligns with the SMBH spin as magnetic flux builds up near the hole as mass accretes.  In this case the wide opening angle of the sheath $\theta_{\rm j,s} \sim 0.3$ results from the large initial misalignment between the spin axis of the SMBH and the orbital plane of the disrupted star (\citealt{StoneLoeb12}).

\subsection{Radio follow-up of TDE flares}
\label{sec:followup} 

Prior to \ubs, TDE flares were detected exclusively via the thermal emission from their transient accretion disks at soft X-ray (e.g., \citealt{Komossa&Greiner99}), UV (e.g., \citealt{Gezari+08}, and optical wavelengths (e.g., \citealt{Chornock+14}, \citealt{Arcavi+14}, \citealt{Holoien:2014}).  The apparent absence of non-thermal X-rays from these events does not in itself rule out the presence of a relativistic jet because the X-ray beaming fraction $f_{\rm b}^{-1} \sim 2/\theta_{\rm j}^{2} \gtrsim 100$ inferred for \ubs~implies that the vast majority of TDEs are viewed off the jet axis.  The presence of a jet should nevertheless manifest itself as a late radio brightening once the jet slows to non-relativistic speeds, allowing the orphan afterglow to become visible (\citealt{Giannios:2011aa}).  In our favoured model for the jet structure of \ubs, which includes an additional wide-angle outflow with a lower, mildly-relativistic initial speed $\Gamma_j \sim 2$, radio emission will become luminous even at early times for most viewing angles.

Follow-up searches have been conducted for radio emission from approximately twenty TDE candidates discovered earlier by their thermal X-ray/optical/UV emission to search for evidence of an off-axis jet (\citealt{vanVelzen:2013aa}, \citealt{Bower:2013aa}).  With three exceptions, most notably IC3599 and RXJ1420+5334 (\citealt{Bower:2013aa}), none show detectable radio emission, in some cases to deep limits (Table \ref{table:data}).  This raises the question of whether a jet with properties similar to that responsible for \ubs~can be ruled out in the majority of TDEs.

\begin{figure}
  \centering
  \includegraphics[width=8.4cm]{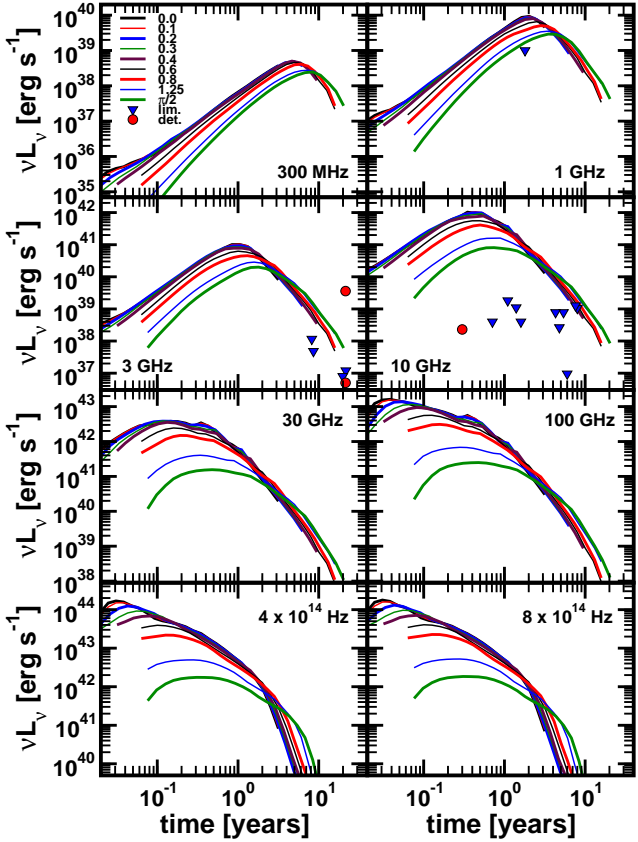}
  \caption{Radio light curves of our fiducial two-component jet model ($\S\ref{sec:2Dlcs}$) for different viewing angles compared to radio detections ({\it red circles}) and upper limits {\it blue triangles}) for thermal TDE flare candidates (Table \ref{table:data}). Legends give the viewing angle in radians. $1$\,GHz radio maps for $0$ and $0.8$ rad viewing angles are shown in \figref{fig:radiomaps}}
  \label{fig:offaxis-2D} 
\end{figure}

Figure \ref{fig:offaxis-2D} compares these radio upper limits ({\it blue triangles}) and detections ({\it red circles}) as a function of time since the flare to our calculation of the off-axis light curves of \ubs~at the frequency closest to that of each observation.   In almost all cases, the model predicted flux exceeds the observational upper limits for all viewing angles, constraining the energy of the relativistic jet in these systems to be typically an order of magnitude less than that in \ubs. The radio detection in RX J1420-5334 is too bright to be consistent with an off-axis J1644+57-like event, while that in IC3599 is somewhat too dim.  Nevertheless, future observations of these events over the coming years can test whether they are fading in the predicted manner.  The early-time radio emission from CSS100217 also appears much too dim to be consistent with an off-axis jet; this supports the interpretation of \citet{Drake:2011aa} that this event was a nuclear Type IIn supernova instead of a TDE. 

\begin{figure}
  \centering
  \includegraphics[width=8.4cm]{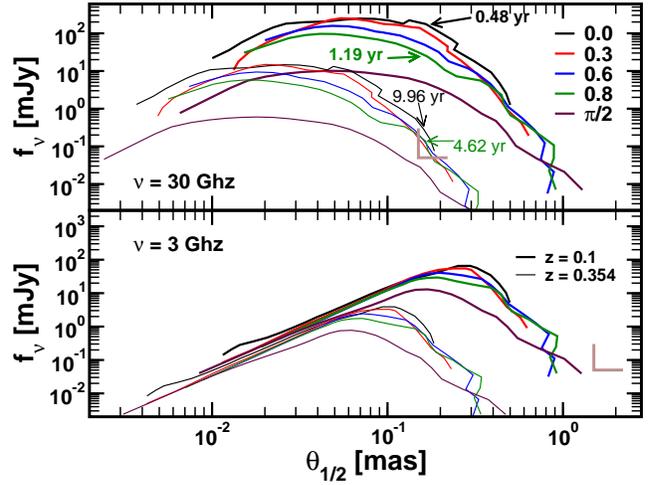}
  \caption{Radio flux observed at $30$\,GHz ({\it upper panel}) and $3$\,GHz ({\it lower panel}) as a function of the angular size $\theta_{1/2}$ of the half-light radius for a source located at redshift $z=0.1$ ({\it \corr{thick} lines}) and $z=0.354$ ({\it \corr{thin} lines}).  Curves are shown for different viewing angles $0$, $0.3$, $0.6$, $0.8$ and $\pi/2$ rad.  The \corr{brown} solid bracket shows an approximation of the flux and angular size required for detection and resolving the source, respectively, with VLBI.  Time marks the point after which the jet is resolved at 30 GHz for a viewing angle of 0 rad (black) and 0.8 rad (green).}
  \label{fig:halflight} 
\end{figure}

The conclusion that most TDE flares are not accompanied by a jet similar to \ubs~raises the question of what conditions are necessary to produce a relativistic jet.  Does jet formation require special `internal' conditions, such as highly super-Eddington accretion or a high black hole spin, or are particular `external' conditions needed to produce detectable emission, such as a low, or high gas density near the SMBH.  If the CNM density were lower, the timescale for jet deceleration and radio rebrightening would increase somewhat, but this delay is unlikely to be sufficiently long to hide the radio emission in these events.  On the other hand, if the gas density were orders of magnitude higher than in \ubs, then the jet could be choked before reaching large radii.  In this case the reverse shock is sufficiently strong to slow the jet to sub-relativistic speeds over just a few weeks, potentially dimming the blast wave emission considerably. 

\begin{figure*}
  \centering
  \begin{tabular}{cc}
    \includegraphics[width=7.5cm]{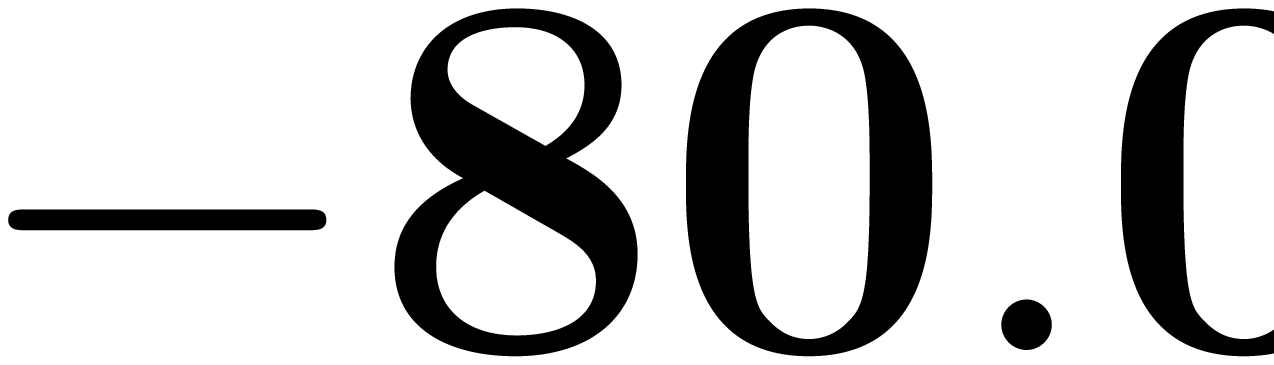} & \includegraphics[width=7.5cm]{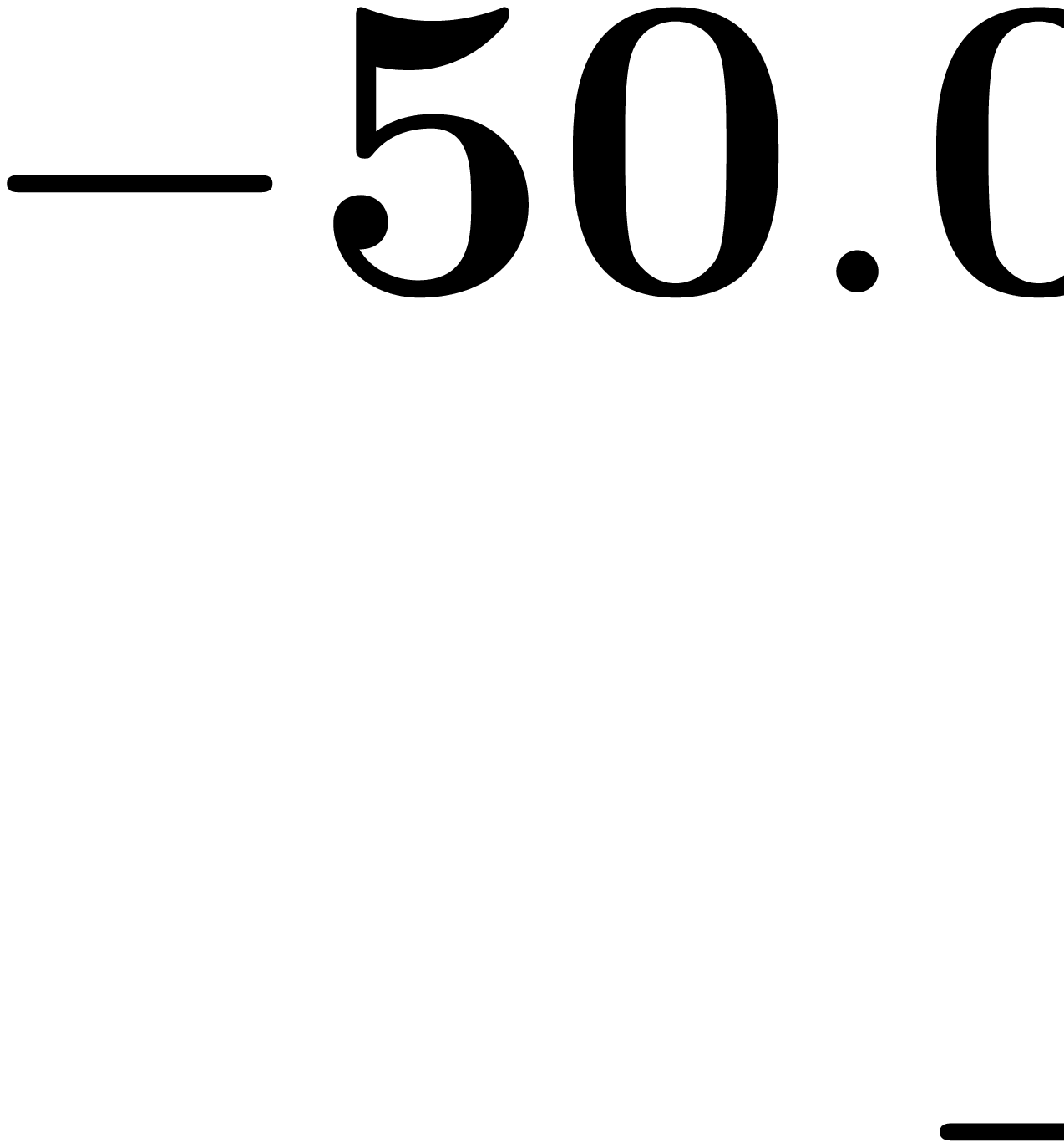} \\
   \includegraphics[width=7.5cm]{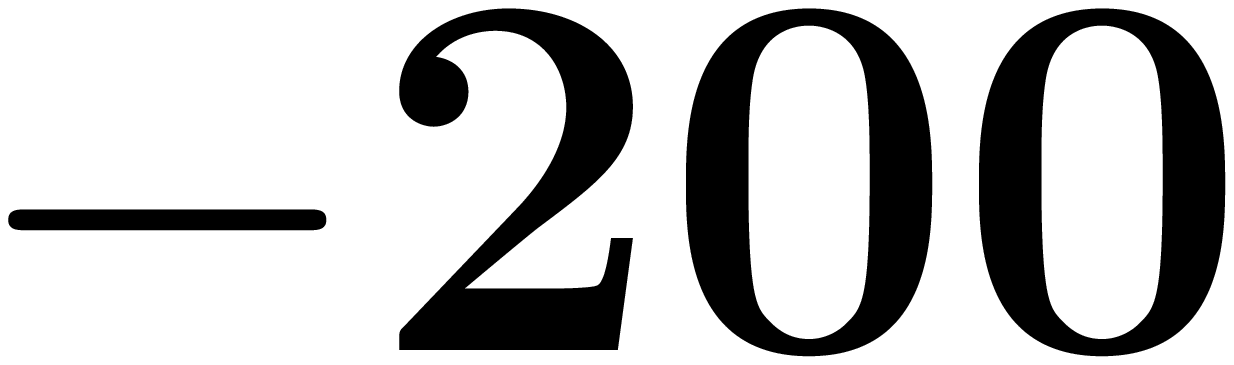} & \includegraphics[width=7.5cm]{figures/s0.8-2.eps}
  \end{tabular}
  \caption{$1$\,GHz radio maps of a two-component jet for $0$ rad and $0.8$ rad viewing angle (left and right panels, respectively). Each panel intensity is normalised separately. Bottom and upper panels correspond to different times after the TDE (see legends). Note that the jet seen at $0.8$ rad is strongly absorbed at early times (upper right panel), while it is almost transparent at $3.41$ years (lower right panel).}
  \label{fig:radiomaps} 
\end{figure*}

Powering a TDE jet via the traditional Blandford-Znajek process may also require the tidal debris to sweep up a large magnetic flux from a fossil accretion disk (\citealt{Tchekhovskoy:2014aa}; \citealt{Kelley+14}).  However, a net large-scale field may not always be necessary to produce a powerful jet (\citealt{Parfrey+14}). Parfrey et al. propose that jets can be driven by magnetic fields amplified locally within the disk, a mechanism which is particularly effective when the disk rotation is retrograde with respect to that of the black hole. Within this interpretation, powerful jets like that responsible for~\ubs~are rare because they are produced only following the disruption of stars with orbits located within the plane perpendicular to, and moving retrograde with respect to, the black hole spin axis.  

\begin{table}
\begin{scriptsize}
\begin{center}
\vspace{0.05 in}\caption{Radio Observations of TDE Candidates}
\label{table:data}
\begin{tabular}{ccccccc}
\hline \hline
\multicolumn{1}{c}{Source} &
\multicolumn{1}{c}{$D_{\rm L}$} & 
\multicolumn{1}{c}{t$^{(a)}$} &
 \multicolumn{1}{c}{$\nu$} &
\multicolumn{1}{c}{$\nu L_{\nu}$} &
\multicolumn{1}{c}{Ref.}
\\
\hline
 &  (Mpc) & (years) & (GHz) & ($10^{36}$ erg s$^{-1}$) \\
\hline 
IC 3599 & 88 & 21.49 & 3 & 5. &  1 \\
RX J1420+5334 & 2970 & 21.49 & 3 & 3600 &  1\\
NGC 5905 & 52 & 21.91 & 3 & $<2.0$ &  1 \\
RX J1624+7554 & 265 & 21.67 & 3 & $<12 $ & 1\\
RX J1242-1119 & 208 & 19.89 & 3 & $ <8$ & 1 \\
SDSS J1323+48 & 365 & 8.61 & 3 & $<48 $ & 1 \\
SDSS J1311-01 & 750 & 8.21 &  3 & $< 115 $ & 1 \\
D1-9 & 1700 & 8.0 & 5 & $<800$ &2\\
D3-13 & 2000 & 7.6 & 5 & $<960$ & 2\\
TDE1 & 645 & 5.4 & 5 & $<120$ & 2\\
D23H-1 & 910 & 4.8 & 5 & $<200$ & 2\\
TDE2 & 1280 & 4.3 & 5 & $<590$ & 2\\
PTF10iya & 1130 & 1.6 & 5 & $<300$ & 2\\
PS1-10jh & 820 & 0.71 & 5 & $<300$ & 2\\
NGC 5905 & 75 & 6.0 & 8.6 & $< 9$ & 3\\
D3$-$13 & 2000 & 1.8  & 1.4 & $< 1000$  &  4\\
TDE2 & 1280 & 1.1 & 8.4 & $< 1650$ &  5\\
CSS100217 & 700 & 0.3 & 7.9 & $230$ &   6\\
SDSS J1201+30 & 700 & 1.4 & 7.9 & $<1000$&  7\\
 \\
  \\
\hline
\hline
\end{tabular}
\end{center}
$^{(a)}$Time since first X-ray or optical detection.   References: (1) \citealt{Bower:2013aa}; (2) \citealt{vanVelzen:2013aa}; (3) \citealt{Bade:1996aa}, \citealt{Komossa:2002aa}; (4) \citealt{Gezari:2008aa}, \citealt{Bower:2011aa}; (5) \citealt{vanVelzen:2011aa}; (6) \citealt{Drake:2011aa}; (7) \citealt{Saxton:2012aa}.  All quoted upper limits are 5$\sigma$.
\end{scriptsize}
\end{table}

\subsection{Prospects for future radio surveys}
\label{sec:survey}

\citet{Frail:2012aa} estimate that \ubs-like events could dominate the number of GHz transients on the sky, with a rate comparable to or exceeding that of GRB orphan afterglows or radio supernovae (cf.~\citealt{Metzger+15}).  Such events thus represent prime targets for upcoming wide-field radio surveys such as ASKAP, MeerKAT, WSRT/Apertif, and, ultimately, the Square Kilometer Array (SKA).

Our calculated off-axis light curves can be used to estimate the rate of off-axis TDEs detected by future surveys.  The rate of jetted TDEs is estimated in two ways.  First, the fact that one nearby event has been detected by {\it Swift} over $\Delta T =$ 10 years of monitoring out to a redshift $z = 0.35$ (co-moving volume $V \approx 11$ Gpc$^{-3}$) suggests a rate of $\sim 1/ (V\Delta T)\sim 0.01\,$Gpc$^{-3}$\,yr$^{-1}$, which given a beaming correction $\sim 100$ for J1644+57 (MGM12) implies a rate $\mathcal{R}_{\rm jTDE} \sim 1$\,Gpc$^{-3}$\,yr$^{-1}$.  Another approach is to take the estimated rate of thermal TDE flares of $\sim 10^{-4}-10^{-5}$\,galaxy$^{-1}$\,yr$^{-1}$ (e.g.~\citealt{VanVelzen&Farrar14}; \citealt{Stone&Metzger15}), which given a local galaxy density $\sim 10^{-2}$\,Mpc$^{-3}$ corresponds to a TDE rate of $\sim 100-1000$\,Gpc$^{-3}$\,yr$^{-1}$.  Assuming that $\lesssim 10\%$ of TDEs produce jets (\citealt{Bower:2013aa}; \citealt{vanVelzen:2013aa}; \figref{fig:offaxis-2D}), this corresponds to a local volumetric rate of $\mathcal{R}_{\rm jTDE}\lesssim 10-100$ Gpc$^{-3}$ yr$^{-1}$ for on-axis jetted TDEs, a factor of $\sim 10-100$ higher than the empirical estimate above. Since the evidence for powerful jets accompanying other TDE candidates are weak, we adopt here the more conservative rate of $\mathcal{R}_{\rm jTDE} \sim 1$\,Gpc$^{-3}$\,yr$^{-1}$  inferred from observations.

A survey with flux sensitivity $F_{\rm lim}$ can detect events to a limiting distance $D_{\rm lim} = D_{0}(F_{0}/F_{\rm lim})^{1/2}$, where $F_{0}$ is the peak flux of the transient at a fiducial distance $D_0$.  If the transient lasts a characteristic duration $\Delta t$, then assuming Euclidean geometry the number of detectable TDEs across a fraction $f_{\rm sky}$ of the sky at any time is approximately given by
\begin{eqnarray}
N_{\rm obs} &=& (4\pi/3)\mathcal{R}_{\rm jTDE}f_{\rm sky}\Delta t D_0^{3}(F_0/F_{\rm lim})^{3/2} \nonumber \\ &\approx& 30\left(\frac{\mathcal{R}_{\rm jTDE}}{\rm 1 Gpc^{-3} yr^{-1}}\right)\left(\frac{f_{\rm sky}}{0.25}\right)\left(\frac{\Delta t}{\rm 3 yr}\right)\left(\frac{F_{\rm lim}}{\rm 0.5\,mJy}\right)^{-3/2},
\label{eq:Nobs}
\end{eqnarray}
where we have used $F_0 = 0.2$\,mJy for $D_0 = 10^{28}\,$cm $\approx 3.2$\,Gpc and $\nu = 1$\,GHz and have taken $\Delta t = 3$\,yr based on the peak flux and duration of our off-axis light curves (\figref{fig:offaxis-2D}).  The flux sensitivity is normalised to a value $F_{\rm lim} \sim 0.5\,$mJy corresponding to the 10$\sigma$ sensitivity of the planned ASKAP VAST-Deep Multi-Field survey, which plans to survey a quarter of the night sky ($f_{\rm sky} = 0.25$) at 1.3 GHz to an rms flux limit $\sigma = 0.05$\,mJy with an approximately yearly cadence (\citealt{Murphy+13}).  We consider that a high detection threshold ($\sim 10\sigma$) will be necessary to overcome the large number of false positive signals resulting from random statistical fluctuations across the large number of observational epochs comprising the survey (\citealt{Metzger+15}).   

Future surveys similar to VAST Deep could thus detect a few tens of off-axis TDE candidates over the course of a few years, depending on the uncertain volumetric rate $\mathcal{R}_{\rm jTDE}$ and how representative \ubs~is of the jetted TDE population.  The number of TDEs detected by the SKA will be substantially higher than the upcoming generation of surveys.  However, the greater sensitivity of the SKA will extend the detection range to high redshifts $z \gtrsim 1$, making the detection rate also sensitive to the uncertain redshift evolution of the TDE rate per galaxy and the supermassive black hole density (e.g.~\citealt{Hopkins+07}; \citealt{Sijacki+14}). 

\subsection{Resolving the jet structure}

Another critical issue is whether the radio emission from \ubs~or TDEs detected by future radios surveys will be resolved by very long baseline interferometry (VLBI) observations (\citealt{Giannios:2011aa}; \citealt{Berger:2012aa}).  Figure~\ref{fig:halflight} shows our calculation of flux of the radio emission from our 2D models as a function of the angular size $\theta_{1/2}$ of the radio emission at 3 and 30 GHz.  In the calculations we ignore the counter jet
that is likely to be present. As a result, we may underestimate the size of the source for misaligned 
jets. The angular size is calculated by first computing the barycenter of the intensity of a radio map (see \figref{fig:radiomaps} for examples of $1$ GHz radio maps), and then determining the diameter of a circle centered on the barycenter containing half of the flux.

Although the angular size increases monotonically, the flux decreases at late times, making it unclear a priori whether the source will remain sufficiently bright to be detected once it becomes large enough to resolve.  Following \citet{Berger:2012aa}, we show for comparison an estimate of the minimum flux required to detect the source, which we take as the 5$\sigma$ flux sensitivity of the EVLA at the closest frequencies to those we consider, 5 and 22 GHz, respectively.  We also assume a best-case angular resolution of 0.15 mas and 1.5 mas at 30 and 3 GHz, respectively.  

For an on-axis source at the distance of \ubs~(z = 0.354; {\it \corr{thin} lines}), resolving the jet appears to be just possible at $\sim 30$ GHz on a timescale of $\sim 10$ years, i.e. within the several years.  This motivates ongoing VLBI observations of \ubs~over the coming years and decade, although the source is unlikely to be resolved by more than a factor of $\sim 2$ in radius before its flux decreases below the detection threshold.  

Prospects are better for a closer source at $z = 0.1$ \corr{thin lines)} for which we predict the source could be resolved at 30 GHz within a year of the TDE.  Since off-axis events are more common, there is a greater chance that one will be observed substantially closer than \ubs~in searches by blind radio surveys ($\S\ref{sec:survey}$).  For every $\sim 40$ off-axis events at $z = 0.354$, approximately one should occur at $z \lesssim 0.1$.  Given our estimated detection rate (eq.~[\ref{eq:Nobs}]), future surveys like ASKAP could over the course of a few years detect a handful of events sufficiently close to resolve the radio structure (upon waiting a few years after the peak flux).  For well-resolved sources, the predicted double-lobed radio morphology could be imaged, and the proper motion of the emission peak could be determined, providing additional constraints on the jet structure.  It may also be possible to distinguish the structured jet model proposed in this work from alternative scenarios.

\subsection{Jet contribution to TDE optical emission}

The bottom panels of \figref{fig:offaxis-2D} show jet synchrotron emission could also contribute to the optical/UV light curves of TDEs.  For on-axis events, the peak optical luminosity can reach $L_{\rm opt} \sim 10^{44}$\,erg\,s$^{-1}$ on timescales of weeks.  Optical emission in \ubs~was obscured by $\sim 8-10$ magnitudes of dust extinction (\citealt{Levan:2011aa}), making this prediction impossible to test.  Unobscured optical emission was detected in J2058+05 in coincidence with the final epochs of BAT detection  (\citealt{Cenko:2012aa}) at luminosities $\sim 10^{44}-10^{45}$\,erg\,s$^{-1}$ for rest-frame frequencies $\nu \sim 1-3\times 10^{15}$\,Hz.  However, the hard spectral slope of this emission, $F_{\nu} \propto \nu^{\beta}$ with $\beta \approx 2$, was consistent with black body emission peaking in the UV rather than non-thermal synchrotron emission.  Such bright blackbody emission characterises other optically- and UV-selected TDEs (e.g., \citealt{Arcavi+14} and references therein) and is believed to be reprocessed UV/soft X-ray emission from the accretion disk (e.g., \citealt{Guillochon+14}), although it is not clear whether all TDEs will possess this emission component (\citealt{Stone&Metzger15}).  

For typical off-axis viewing angles ($\theta_{\rm obs} \gtrsim 0.8$), we predict peak optical luminosities from the TDE jet of $L_{\rm opt} \sim 10^{42}-10^{43}$\,erg\,s$^{-1}$, similar to those of supernovae and lasting for several months.  The chief distinguishing features of TDE jet transients are (1) flat, non-thermal spectra and (2) unusually long-durations of several months or longer.  Given that the beaming-corrected rates of jetted TDEs is at least a factor $\sim 10^{4}$ times lower than the rate of Type Ia supernovae, it is not surprising that no jetted TDEs have yet been detected by optical surveys.

\section{Conclusions}\label{sec:concl}

This paper explores the evolution and structure of relativistic jets created following the tidal disruption of stars by massive black holes in the nuclei of galaxies.  Our work is motivated by the discovery of \ubs, the prototypical ``jetted TDE" that has been the target of extensive and ongoing radio follow-up observations. The radio afterglow of \ubs~contains valuable information on the jet acceleration, emission, and structure.  It also provides a unique probe of the radial density structure of the CNM surrounding otherwise quiescent black holes.  

The unexpected radio rebrightening of \ubs~defied the expectations of simple jet models usually applied to GRB afterglows, motivating a proliferation of proposed explanations in the literature.  We have performed a comprehensive exploration of these possibilities by means of a large number of hydrodynamical simulations of the jet-CNM interaction in order to find a jet model that can best reproduce the multi-frequency radio data of \ubs.

After experimenting with  a large number of one-dimensional models (most of which are not shown here for brevity) with varying jet speed, opening angle and CNM density and profile, we conclude that no homogeneous, ``top hat'' jet model adequately reproduces the entire light curve evolution.  Top hat jets can be made to fit either the early radio bump or the late-time plateau, but not both of these prominent features simultaneously. We instead favor a jet with a hybrid angular structure, comprised of a narrow, fast ($\Gamma\sim 10$)
core and and a wide slow ($\Gamma\sim 2$) sheath, each containing a comparable amount of energy.  Our 2D axisymmetric simulations verify that such a model works well in explaining the observations.  Such a geometry could result from the radial structure of a super-Eddington accretion flow or as the result of jet precession prior to alignment of the jet with the black hole spin axis.

Radio observations extending from days after the gamma-ray trigger to years or longer probe both the initial relativistic jet stages of jet-CNM interactions as well the late  mildly and non-relativistic evolution of the blast. Simple analytical treatments that assume self-similar, spherically expanding blast waves
cannot be applied to the totality of radio data. Models that assume a spherically symmetric situation overpredict the resulting emission several months after the TDE by one order of magnitude or more.  Models that assume conical jet motion but take into account the finite jet opening angle $\theta_{\rm j}$ into the emission calculation give much more consistent results in comparison to the fully 2D simulations (though discrepancies of order of $\sim 3$ remain).

Among the models we have considered in this paper, the ones which are closer to the observational data of \ubs~possesses a combined kinetic energy $\simeq 5\times 10^{53}$ erg  in mildly- and ultra-relativistic ejecta.  Although this precise value is somewhat degenerate with our assumed values of the shock microphysical parameters, a large energy $\gtrsim 10^{53}$erg is required of any model that can successfully account for the late radio bump (see also \citealt{Berger:2012aa}; \citealt{Barniol:2013aa}).  This is close to the gravitational energy released by the accretion of a solar mass star, implying that the jet launching mechanism in some TDEs must be very efficient. \corr{We note that our model underpredicts the early-time emission at frequencies $\simless 10$ Ghz, but this could be corrected for by adopting somewhat non-standard values for the microphysical parameters (Appendix~\ref{app:zeta_e}).}

The dynamical evolution of 2D axisymmetric jets flanked by a slower sheath is significantly different than the evolution of (single component) ``top hat'' jet models. The former are much more stable against Kelvin-Helmhotz instabilities and display a faster jet propagation speed during the first months after the TDE. Furthermore, spine-sheath jets, as the one considered here, remain well collimated for much longer time than equivalent ``top hat'' jets. Thus, the aspect ratio of two-component jets is more prolate than in one-component jets (the former develop rather spherical cavities on time-scales of months). If the jet cavity could be resolved using VLBI observations, its aspect ratio could provide us clues about the jetted nature of the event.

The opening angle that we infer for the fast jet core $\theta_j\sim 0.1$ results in an X-ray beaming fraction $f_{\rm b}^{-1} \sim 100$.  The resulting beaming-corrected rate of jetted TDEs, $\mathcal{R}_{\rm jTDE} \sim 1$ Gpc$^{-3}$ yr$^{-1}$, is less than one percent of the TDE rate estimated by other means (see, e.g., \citealt{Stone&Metzger15}).  This finding is consistent with the lack of evidence for off-axis jets in TDEs detected by their thermal accretion disk emission (Fig.~\ref{fig:offaxis-2D}).  Special conditions appear necessary for the creation of a powerful jet following TDEs.

Off-axis jetted TDEs represent a promising source for upcoming radio surveys.  We present self-consistent calculation of the off-axis radio light curves of jetted TDEs, based on a 2D model that provides a reasonable fit to the light curve of the on-axis event \ubs.  Planned surveys such as ASKAP VAST could detect tens of jetted TDEs. The angular size of the afterglow of \ubs\, at $\sim 20-30$ GHz exceeds the estimated VLBI resolution limit of 0.2 mas on a timescale of $\sim$ 10 years, at which point the flux density is sufficiently high to be detected.   Closer, off-axis TDEs detected without X-ray triggers by radio surveys are well-resolved on timescales of a year or less, allowing for a detailed study of the jet structure.

\section*{Acknowledgments}
We thank E.~Berger, B.~Zauderer, and A.~Soderberg for helpful converzations.  PM and MAA acknowledge the support from the European Research Council (grant CAMAP-259276), and the partial support of grants AYA2013-40979-P and PROMETEO-II-2014-069.  DG acknowledges support from the NASA grant NNX13AP13G.  BDM gratefully acknowledges support from the NSF grant AST-1410950 and the Alfred P. Sloan Foundation. We thankfully acknowledge the computer resources, technical expertise and assistance provided by the "Centre de C\`alcul de la Universitat de Val\`encia" through the use of {\emph{Llu\'{\i}s Vives}} cluster and {\emph{Tirant}}, the local node of the Spanish Supercomputation Network.

\appendix

\section{Technical details and tests}
\label{app:conv}

In this section we briefly discuss some technical details of the light curve calculation in \verb'SPEV'. As was already discussed in Sections~\ref{sec:1Ddetails} and \ref{sec:2Ddetails}, the results of a numerical hydrodynamics simulation are periodically stored so that at the end of the simulation a collection of snapshots can be used to produce the synthetic observations. An important parameter is the frequency $1 / \Delta T_{\rm snap}$ at which the snapshots are being stored. We chose $\Delta T_{\rm snap} = 8.34\times 10^4$ s because it provides a good trade-off between the space-time coverage of the jet evolution and the technical limitations\footnote{We note that the choice of $\Delta T_{\rm snap}$ was in large part constrained by the fact that size of the intermediate files and the computational time depend as $\Delta T_{\rm snap}^{-2}$ \citep{Aloy:2013az}.}. In \figref{fig:app_freq} we show that numerical light curves computed by \verb'SPEV' for $\Delta T_{\rm snap} = 8.34\times 10^4$ s are very close to those computed using $\Delta T_{\rm snap} = 4.17\times 10^4$ and $1.67\times 10^5$ s.

\begin{figure}
  \centering
  \includegraphics[width=8.4cm]{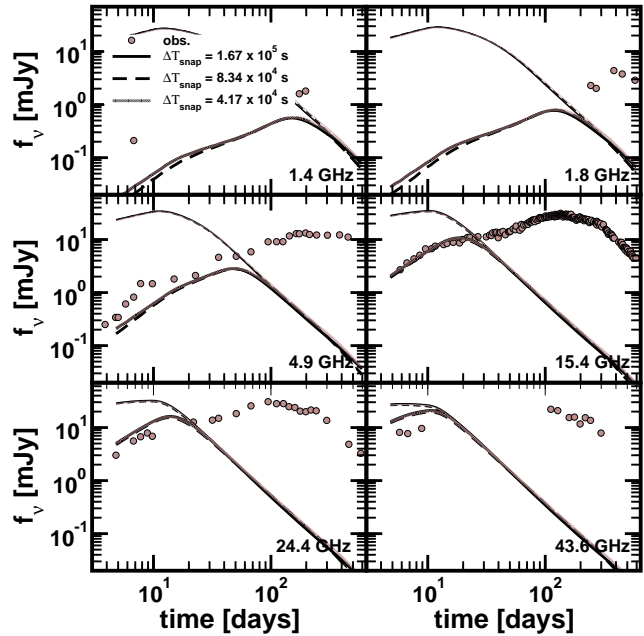}
  \caption{Same as \figref{fig:FNjet}, but showing the result of the models computed with lower and higher snapshot frequency. The model with $\Delta T_{\rm snap} = 8.3 \times 10^4$ s (see text for details) corresponds to the $n(r)\propto r^{-3/2}$ model in \figref{fig:FNjet}. \corr{The differences in the light curves between the various angular resolutions are so small that all models seem to lie on top of each other.}}
  \label{fig:app_freq} 
\end{figure}

Next we study the influence of the angular resolution on the light curves. For 1D models in this paper we use $510$ angular zones per radian, i.e. $N_\theta = 51$ zones in a narrow jet. \figref{fig:app_ntheta} shows that that value is adequate since varying $N_\theta$ by a factor of a few does not change our results.

\begin{figure}
  \centering
  \includegraphics[width=8.4cm]{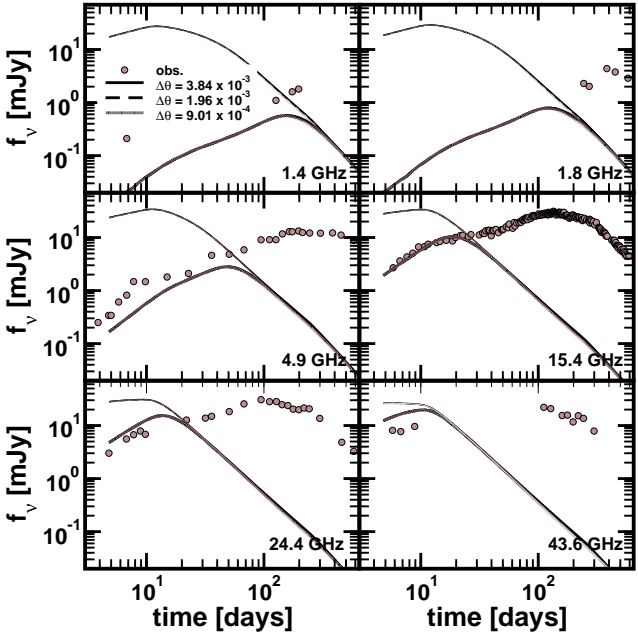}
  \caption{Same as \figref{fig:FNjet}, but showing the result of the models computed with lower and higher angular resolution. The model with $\Delta\theta = 1.96\times 10^{-3}$ rad corresponds to the $n(r)\propto r^{-3/2}$ model in \figref{fig:FNjet}.}
  \label{fig:app_ntheta} 
\end{figure}

Finally, we study the influence of the discretization in non-thermal electron momentum space (see Sec. 3.2 of \citealt{Mimica:2009aa} for more details). The electron Lorentz factors are initially distributed logarithmically in $N_b$ bins,
\begin{equation}
  \gamma_i = \gamma_{\rm min}\left(\dsfrac{\gamma_{\rm max}}{\gamma_{\rm min}}\right)^{(i-1)/(N_b-1)}\, ,
\end{equation}
where $\gamma_{\rm min}$ and $\gamma_{\rm max}$ are initial lower and upper cutoffs. \figref{fig:app_bins} shows the comparison of the light curves for $N_b = 32$, $64$ and $128$. We note that the size of the intermediate files scales as $\sim N_b$ and chose $N_b = 64$ as a good compromise between the file size and light curve accuracy. 
\begin{figure}
  \centering
  \includegraphics[width=8.4cm]{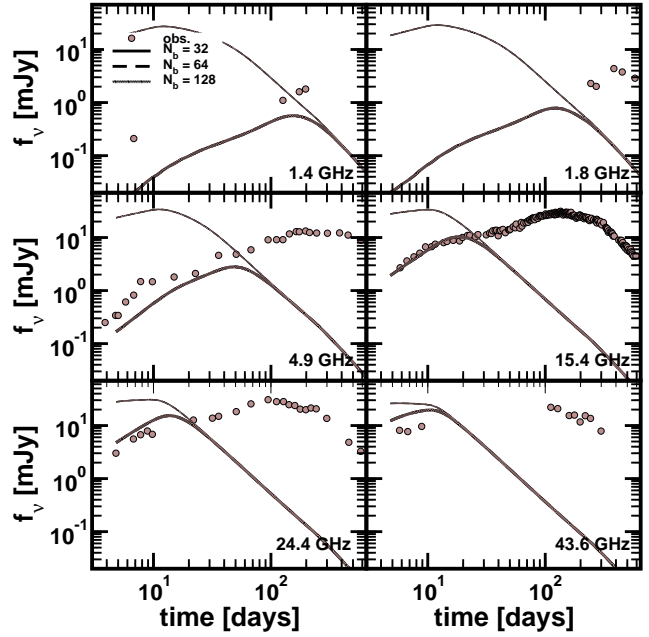}
  \caption{Same as \figref{fig:FNjet}, but showing the result of the models computed with different number of bins $N_b$. The model with $N_b = 64$ corresponds to the $n(r)\propto r^{-3/2}$ model in \figref{fig:FNjet}.}
  \label{fig:app_bins} 
\end{figure}

\section{Two-component light curves with $\zeta_e = 0.1$}
\label{app:zeta_e}

In this section we study the case when the number of electrons accelerated to relativistic energies is a fraction $\zeta_e$ of the electrons crossing the shock (see equation~14 in \corr{Section~3.3} of \citealt{Mimica:2012aa} for the definition and discussion of $\zeta_e$). We set $\zeta_e=0.1$ and compute light curves for our reference 1D two-component model (Section~\ref{sec:1D2Z}) and 2D two-component model (Section~\ref{sec:2D}), \corr{and show the results in Fig.~\ref{fig:zeta_e}. It turns out that reducing the value of $\zeta_e$ below the standard one reduces the optical depth, thus increasing the flux at all times, but especially at early times and at low frequencies. In other words, in the optically thin regime the change in $\zeta_e$ by one order of magnitude has very little impact on the light-curve, while in the optically thick regime the flux is roughly inversely proportional to $\zeta_e$.  The peak flux and the time when it happens does not strongly depend on $\zeta_e$.  In the 1D models, the late time differences between the cases $\zeta_e=0.1$ and $\zeta_e=1$ are small. 
} 

\begin{figure}
  \centering
  \includegraphics[width=8.4cm]{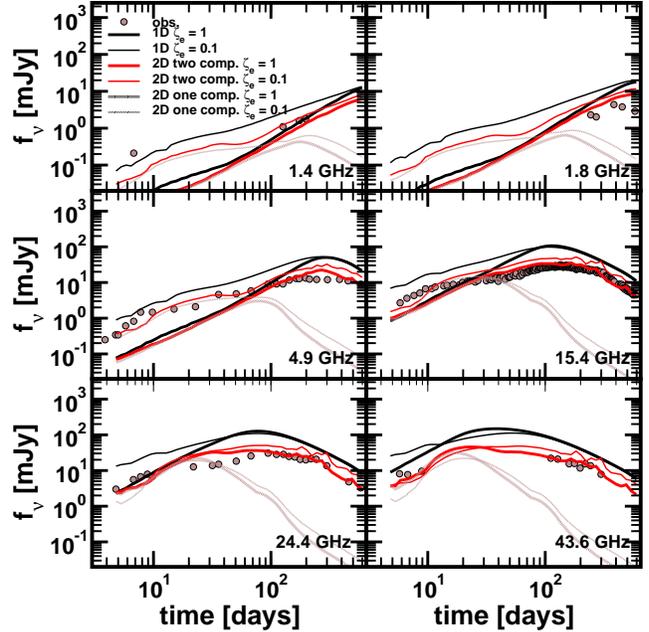}
  \caption{Same as \figref{fig:FNjet}, but showing the light curves produced by 1D (full lines) and 2D two-component (red) and one-component (grey lines) jet simulations. We use $\epsilon_e = 0.1$, $\epsilon_B = 0.002$, and show the light curves for $\zeta_e = 1$ (thick lines) and $\zeta_e = 0.1$ (thin lines). In all simulations $E_{\rm iso} = 4\times 10^{54}$ erg. }
  \label{fig:zeta_e} 
\end{figure}

\corr{
Figure~\ref{fig:zeta_e} also shows that with a smaller value of $\zeta_e$ the 1D model tends to overpredict the early time flux at all frequencies except at 1.4\,GHz. However, the decrease in optical thickness due to lower $\zeta_e$ is not enough for the 2D models to offset the flux deficit in our synthetic multidimensional models at $t\lesssim 20$\,days. We note that this deficit happens both in the 2D two component jet model and in the 2D fast-narrow jet model, but the latter still exhibits orders of magnitude flux deficit at late times (thus eliminating a single component model as a suitable explanation for the observed phenomenology). This probably means that the emission originating at the reverse shock has to be taken into account in those models where the optical depth at early times is smaller (i.e., when $\zeta_e\ll 1$). 

In light of the results shown in Fig.~\ref{fig:zeta_e}, it is difficult to assess by eye which value of $\zeta_e$ is optimal to best explain the observations. Thus, we have performed a simple error analysis by computing the vertical distance from the data to the model for each observational frequency $\nu$, defined as $e_i(\nu) =  {\rm data}_i(\nu) - {\rm model}_i(\nu)$, where the index $i$ runs over observational times and ${\rm model}_i(\nu)$ is obtained by  interpolation between the two model points computed at times that bracket any given observational data. We define the total error at a given observational frequency as
\begin{equation}
D_\nu:=\frac{1}{N_\nu}\sum_{i=1}^{N_\nu}  \log [\cosh(e_i(\nu))],
\label{eq:errors}
\end{equation}
where $N_\nu$ is the number of observational data points at each frequency $\nu$.
 
As can be observed in Tab.~\ref{table:errors}, in spite of the fact that the case $\zeta_e=0.1$ clearly improves the representation of the observational data for our reference 2D two component model at $\nu=1.4\,$GHz, there is no quantitative reason to choose this particular value instead of the standard $\zeta_e=1$. At the lowest frequency ($\nu=1.4\,$GHz) the total error is about two times smaller for $\zeta_e=0.1$ than for $\zeta_e=1$, but this difference is compensated by the better behavior of the model with $\zeta_e=1$ at frequencies $\nu\geq 1.8$\,GHz. This quantitative analysis reinforces our choice of fiducial parameters (particularly the value of $\zeta_e$) to show that our models may accomodate the observations with moderate and sound variations of the microphysical parameters.  Taken all together, our models may actually {\em fit} the observations with a suitable combination of a value of $\zeta_e$ in the range $[0.1,1]$ and the contribution of the reverse shock at early times if the optical depth is reduced. 
\begin{table}
\begin{center}
\vspace{0.05 in}\caption{Difference $D_\nu$ between the observations and our models.}
\label{table:errors}
\begin{tabular}{ccccc}
\hline \hline\\*[-0.5pt]
\multicolumn{1}{c}{$\nu$} &
\multicolumn{2}{c}{1D} & 
\multicolumn{2}{c}{2D} 
\\
\multicolumn{1}{c}{(GHz)} &
\multicolumn{1}{c}{$\zeta_e=1$} & 
\multicolumn{1}{c}{$\zeta_e=0.1$} & 
\multicolumn{1}{c}{$\zeta_e=1$} & 
\multicolumn{1}{c}{$\zeta_e=0.1$} 
\\
\hline 
1.4   &   0.02  &       1.02  &     0.09  &    0.04\\
1.8   & 6.73 &       10.20  &       2.12      &  3.93\\
4.9   & 9.51  &      14.36 &       1.73    &    4.23\\
15.4 &  40.37 &       40.84&        3.54   &     11.46\\
24.4 & 36.89  &      38.23 &        4.42     &   10.63\\
43.6 & 25.81  &      28.68 &       1.52    &    8.40\\
\hline
\hline
\end{tabular}
\end{center}
\corr{
In the second and third column we show the total error (Eq.~\ref{eq:errors}) between the 1D jet models computed with parameters $\zeta_e=0.1, 1$. In the fourth and fifth columns, the same error is computed for the reference 2D two component jet model. The light curves corresponding to all these models are plotted in Fig.~\ref{fig:zeta_e}.}
\end{table}
}


\bibliographystyle{mn2e}
\bibliography{mimica.bib}

\end{document}